\documentclass[conference]{IEEEtran}

\IEEEoverridecommandlockouts

\usepackage{amsmath,amssymb,amsfonts,amsthm}
\usepackage{graphicx}
\usepackage{subfigure}
\usepackage{textcomp}
\usepackage{xcolor}
\usepackage{multirow}
\usepackage{enumitem}
\usepackage{threeparttable}
\usepackage{float}
\usepackage[ruled,vlined,linesnumbered]{algorithm2e}
\usepackage{tabularray}
\usepackage{makecell}
\usepackage[hidelinks]{hyperref}
\usepackage{cleveref}
\makeatletter
\let\icde@nl@sethref\algocf@nl@sethref
\renewcommand{\algocf@nl@sethref}[1]{%
  \icde@nl@sethref{#1}%
  \gdef\@currentcounter{AlgoLine}}
\makeatother
\crefname{figure}{Fig.}{Figs.}
\Crefname{figure}{Fig.}{Figs.}

\newtheorem{definition}{Definition}
\newtheorem{lemma}{Lemma}
\newtheorem{problem}{Problem}

\makeatletter
\renewcommand{\paragraph}[1]{%
  \par\vspace{0.75ex plus 0.2ex minus 0.1ex}%
  \noindent{\normalfont\normalsize\bfseries #1.}\hspace{0.5em}\ignorespaces}
\makeatother

\AtBeginDocument{%
}

\begin{document}

\title{Scalable Algorithm for\\ Dynamic Quasi-clique Detection}

\author{%
\IEEEauthorblockN{Jingbang Chen\IEEEauthorrefmark{1}}
\IEEEauthorblockA{\textit{CUHK-Shenzhen \& SLAI}\\
chenjb@cuhk.edu.cn}
\and
\IEEEauthorblockN{Weinuo Li\IEEEauthorrefmark{1}}
\IEEEauthorblockA{\textit{Zhejiang University}\\
liweinuo@zju.edu.cn}
\and
\IEEEauthorblockN{Yingli Zhou\IEEEauthorrefmark{2}}
\IEEEauthorblockA{\textit{CUHK-Shenzhen}\\
yinglizhou@link.cuhk.edu.cn}
\and
\IEEEauthorblockN{Hao Wu}
\IEEEauthorblockA{\textit{Zhejiang University}\\
haowu2318@gmail.com}
\and[\hfill\mbox{}\par\mbox{}\hfill]
\IEEEauthorblockN{Can Wang}
\IEEEauthorblockA{\textit{Zhejiang University}\\
wcan@zju.edu.cn}
\and
\IEEEauthorblockN{Yixiang Fang}
\IEEEauthorblockA{\textit{CUHK-Shenzhen}\\
fangyixiang@cuhk.edu.cn}
\and
\IEEEauthorblockN{Chenhao Ma\IEEEauthorrefmark{2}}
\IEEEauthorblockA{\textit{CUHK-Shenzhen}\\
machenhao@cuhk.edu.cn}
\thanks{\IEEEauthorrefmark{1}Jingbang Chen and Weinuo Li contributed equally. \IEEEauthorrefmark{2}Yingli Zhou and Chenhao Ma are the corresponding authors.}%
}

\maketitle

\begin{abstract}
Identifying dense subgraphs known as quasi-cliques is pivotal in numerous graph mining tasks across domains such as social networks, biology, and e-commerce. While prior work has developed efficient algorithms for quasi-clique detection in static graphs, real-world networks are inherently dynamic, where edges appear and disappear continuously. This renders static methods inefficient and ill-suited for real-time analysis.
In this paper, we initiate the study of the Dynamic Maximum Quasi-Clique Problem (DMQCP), which aims to maintain and update the largest quasi-clique in a graph under streaming graph updates. We propose \texttt{DMI}, a novel MinHash-based dynamic framework that supports fast, high-quality approximate maintenance of quasi-cliques. \texttt{DMI} leverages two update-efficient hashing schemes, i.e., $l$-buffered $k$-MinHash and Bottom-$k$ MinHash, to maintain candidate quasi-cliques incrementally. To ensure robustness and reduce bias, we further design a batch reconstruction strategy to periodically rebuild the candidate set, guaranteeing both stability and adaptability under frequent updates.
Extensive experiments on real-world and synthetic datasets show that \texttt{DMI} achieves up to four orders of magnitude speedup over static baselines, while preserving solution quality. As a side product, we also propose a framework \texttt{NSF} that primarily uses the neighbor-search technique to maintain quasi-clique candidates while edge updating. This work establishes the first efficient algorithmic framework for dynamic quasi-clique extraction, enabling scalable and real-time dense subgraph mining in evolving networks.
\end{abstract}

\begin{IEEEkeywords}
quasi-clique, dynamic graph, MinHash, dense subgraph mining
\end{IEEEkeywords}

\section{INTRODUCTION}
%dense subgraph extraction
Dense subgraph extraction is a fundamental problem in graph mining and network analysis, with the aim of identifying vertex subsets that exhibit a high level of internal connectivity. In complex networks in the real world, such as social, communication, biological, and financial networks, dense regions often correspond to cohesive communities, collaborative groups, or functionally related entities. Formally, the goal is to extract subgraphs that optimize a density measure, such as edge density \cite{zhou2024depth}, or clique density \cite{zhou2025efficient,zhou2024counting}. Classical formulations include the densest subgraph problem \cite{ma2022convex,zhou2024depth}, $k$-core decomposition \cite{batagelj2003m,seidman1983network}, and $k$-clique densest subgraph detection \cite{tsourakakis2015k,fang2022densest,zhou2024counting,zhou2025efficient}. Each formulation offers a different balance between computational tractability and expressiveness in capturing dense patterns. Due to its interpretability and wide applicability, dense subgraph discovery has been extensively studied and applied in various fields, including social network analysis \cite{fang2020survey}, graph compression \cite{ma2021efficient}, fraud/anomaly detection \cite{beutel2013copycatch,hooi2016fraudar,gibson2005discovering}, and identification of regulatory motifs in genomic DNA \cite{jiang2006network}.

Among different formulations, clique is one of the most classic dense subgraph models, where every pair of vertices shares an edge. The maximum clique problem, which seeks the largest such subgraph, is a cornerstone of combinatorial optimization and is well known to be NP-hard~\cite{pattillo2013maximum}. Despite its computational complexity, clique extraction remains highly relevant because cliques embody the strongest possible notion of cohesion within a network. However, while cliques capture complete connectivity, they are often too restrictive in the presence of noise or incomplete data. Quasi-clique extraction relaxes this strict requirement by allowing a limited number of missing edges, thereby generalizing the notion of high density.

\begin{figure}[t]
    \centering
    \includegraphics[width=0.82\columnwidth]{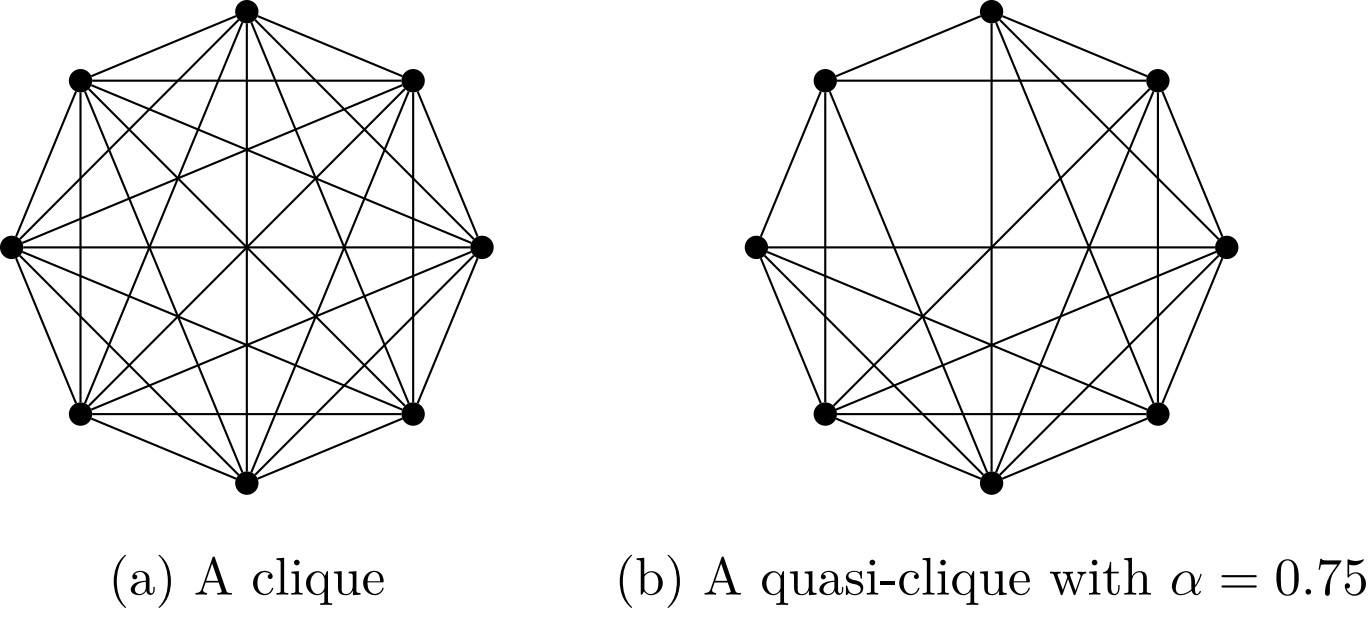}
    \caption{Illustrating clique and quasi-clique.}
    \label{fig:sample}
\end{figure}

An $\alpha$-quasi-clique denotes a subgraph such that the number of edges is at least $\alpha \in (0,1]$ times the number of edges in the clique of the same size. To illustrate, we give a clique and a quasi-clique of 8 vertices in \Cref{fig:sample}. For full connectivity, there are 28 edges in total. In (b), only 21 edges are present, giving an edge-density of $21/28=0.75$.
Quasi-cliques offer a more flexible formulation by requiring the subgraph to be nearly fully connected according to an edge-density threshold $\alpha$. As a result, this relaxation enables us to better capture real-world structures that deviate slightly from perfect connectivity but still exhibit strong internal cohesion. Quasi-clique extraction has already been applied to identify functional modules in protein--protein interaction networks \cite{tsourakakis2013denser,dai2022fast}, uncover approximate communities on social graphs \cite{pang2024similarity}, and detect dense but incomplete collaboration groups \cite{ribeiro2019exact,dai2023maximal}. 

Recent research continues to develop scalable algorithms to extract large quasi-cliques~\cite{pang2024similarity,chen2021nuqclq,ribeiro2019exact,veremyev2016exact,konar2020mining} from a static graph, addressing the maximum quasi-clique problem (MQCP), which finds the largest $\alpha$-quasi-clique in a graph. However, real-world networks are rarely static as they evolve continuously, with vertices and edges being added and removed over time. Examples include social networks where interactions fluctuate, communication networks where connections appear and disappear, and biological networks where relationships among entities change with conditions. Since dense substructures evolve along with the network itself, this motivates us to consider the quasi-clique extraction in a dynamic setting. Unlike static quasi-clique extraction, which analyzes a single snapshot, the dynamic setting aims to track, maintain, and update dense subgraphs efficiently as the graph changes. This is essential to capture the temporal dynamics of communities, enabling real-time analysis. Despite its usefulness, \textbf{there is no prior work studying the dynamic quasi-clique extraction problem}, which naturally presents challenges in balancing accuracy, scalability, and update efficiency.

To bridge this gap, we initiate the study of the \textit{Dynamic Maximum Quasi-clique Problem (DMQCP)} that answers multiple queries of the largest $\alpha$-quasi-clique after inserting or deleting an edge each time. We extend the prior MQCP framework~\cite{pang2024similarity} to solve the dynamic problem. Specifically, we utilize two different MinHash algorithms that support updates: $l$-buffered $k$-MinHash~\cite{10.1145/3701551.3703491} and Bottom $k$-MinHash~\cite{thorup2013bottom}. Beyond keeping the largest one, we maintain a list of candidate quasi-cliques. When an edge is inserted or deleted, we first update the hashing signature and degrees for each relevant vertex. Then, we check if new quasi-cliques need to be added to the list or if any existing ones must be deleted or replaced. To avoid bias, we also utilize the batch reconstruction strategy, rebuilding the entire list after enough rounds of updating. As a result, we give a new framework named \texttt{DMI} and evaluated it on both real-world and synthetic data. Our contributions are summarized as follows:
\begin{itemize}[leftmargin=*]
    \item We design a novel framework named \texttt{DMI} that produces high-quality approximate solutions for the maximum quasi-clique problem under the dynamic setting. 
    \item Two different MinHash methods are applied to \texttt{DMI}, and we give both theoretical and empirical evaluations on their effectiveness.
     \item We also propose a framework named \texttt{NSF}, that primarily uses the neighbor-search technique to directly maintain quasi-cliques.
    \item Extensive experiments\footnote{Code and experiment scripts are available at \href{https://github.com/freewh2005/DMI}{\nolinkurl{https://github.com/freewh2005/DMI}}.} on both real-world and synthetic data show that our algorithm is up to four orders of magnitude faster than baselines while extracting high-quality quasi-cliques.
    % \item Additional experiments are conducted to compare different parameters, providing a more comprehensive understanding of the empirical performance.
\end{itemize}

\paragraph{Outline} The rest of the paper is organized as follows. We review the related work in \Cref{sec:relatedwork}. \Cref{sec:prelim} gives the necessary notations, introducing the dynamic maximum quasi-clique problem and the MinHash technique. In \Cref{sec:algo}, we formally propose our algorithm and give a theoretical analysis. \Cref{sec:neighbor} discusses the neighbor-search-based framework. Empirical evaluations are given in \Cref{sec:experiment}. We conclude our paper in \Cref{sec:conclude}.
\section{RELATED WORKS}
\label{sec:relatedwork}
In this section, we review two representative classes of clique-related problems, including clique problems and quasi-clique problems, on both static and dynamic graphs.

\paragraph{Clique problem} The clique problem is a fundamental topic in graph theory and data mining, encompassing several classic variants: (1) clique counting and enumeration, which aims to identify and count all cliques of size $k$ within a graph \cite{cui2013online,finocchi2015clique,jain2020power,li2020ordering,danisch2018listing,ye2022lightning}; (2) maximal clique enumeration, which seeks all cliques that are not strictly contained in any larger clique \cite{dai2022fast,tomita2006worst,eppstein2013listing,chang2013fast,makino2004new,cheng2011finding,deng2024accelerating,zhou2024efficient}; and (3) maximum clique detection, which aims to identify the clique with the maximum number of vertices in a graph \cite{chang2019efficient,lu2017finding}.
These problems have been further extended to various graph settings.
In bipartite graphs, the analogous structure is the bi-clique~\cite{chen2022efficient,lyu2020maximum,chen2025scalable,mang2025}.
In directed graphs, the notion of cliques generalizes to directed cliques~\cite{conte2021maximal}.
Recent studies have also explored dynamic clique problems, where the underlying graph evolves over time.
Dynamic algorithms for $k$-clique counting~\cite{dhulipala2021parallel} aim to efficiently update clique statistics under edge insertions and deletions without recomputing from scratch.
Similarly, several incremental and fully dynamic frameworks~\cite{yang2021dynamic,das2019incremental} have been developed to maintain and update maximal or maximum cliques as the graph changes.

\paragraph{Quasi-clique problem} Due to the strict connectivity of cliques and the sparsity or noise in real-world data, researchers proposed quasi-cliques---relaxed clique definitions. These can be broadly classified into two types: (1) vertex-based and (2) edge-based. (1) Vertex-based quasi-cliques. This group includes the $k$-plex \cite{balasundaram2011clique,zhou2021improving} and the $\delta$-clique \cite{abello2002massive,yu2023fast}.
In a $k$-plex of size $n$, each vertex may miss edges to at most $k-1$ others, i.e., its degree is at least $n-k$.
Similarly, in a $\delta$-clique, each vertex must connect to at least $\delta \cdot (n-1)$ neighbors, where $\delta \in (0,1]$ defines the minimum connectivity ratio.
(2) Edge-based quasi-cliques. A $k$-defective clique \cite{dai2023maximal,chang2023efficient} allows up to $k$ missing edges within a vertex set $S$, i.e., it contains at least $\binom{|S|}{2} - k$ edges.
An $\alpha$-quasi-clique \cite{tsourakakis2013denser,pang2024similarity} requires the edge density to be at least $\alpha$, meaning it includes at least $\alpha \cdot \binom{|S|}{2}$ edges.
Both maximal and maximum versions of these problems have been extensively studied.
The quasi-clique problems have also been extended to other types of graphs, such as bipartite graphs and directed graphs \cite{guo2022maximal,tsourakakis2013denser,chang2023efficient,pang2024similarity}.
In this work, we focus on the maximum $\alpha$-quasi-clique problem under dynamic graph settings.
To the best of our knowledge, no prior work has effectively addressed the maintenance or incremental discovery of $\alpha$-quasi-cliques in dynamic graphs.

\section{PRELIMINARIES}
\label{sec:prelim}
\subsection{Notation}
We consider an unweighted and undirected graph $G(V,E)$, where $V(G)$ and $E(G)$ denote the vertex set and edge set of $G$, respectively. We denote the numbers of vertices and edges in $G$ by $n=|V(G)|$ and $m=|E(G)|$. For any vertex $u$, we use $N(u)$ to represent the set of nodes that are neighbors of $u$ and $u$ itself. In addition, $N(u,S)$ denotes the neighborhood of $u$ in $S$, i.e., $N(u,S) = N(u) \cap S$. The degree of $u$ is defined as the number of $u$'s neighbors, denoted as $d(u)$. Similarly, $d_S(u)$ denotes the number of neighbors of $u$ in $S$. We denote $d_{max}$ as $\max_{u\in V}d(u)$. Given a subset of vertices $S \subseteq V$, denote $E(S)$ as the subset of $E$ containing edges only between the vertices in $S$, i.e., $E(S) = E \cap (S \times S)$.

We are now ready to define the edge density.
\begin{definition}[Density~\cite{ma2022convex,chen2021nuqclq,pang2024similarity}]
\label{def:density}
Given a graph $G = (V, E)$ and its subgraph $G_S = (S, E(S))$ induced by $S$ with $|S|\geq2$, its edge-density is defined as:
\begin{equation}
\delta(S) = \frac{|E(S)|}{\binom{|S|}{2}}
\end{equation}
\end{definition}

A clique is a subset of vertices such that there exists an edge between every pair of vertices. Therefore, $\delta(S) = 1$ implies $S$ is a clique. Given a parameter $\alpha \in (0,1]$, a subgraph $G_S$ is said to be an $\alpha$-quasi-clique if $\delta(S) \ge \alpha$, indicating that the number of its edges is at least $\alpha \cdot \binom{|S|}{2}$.

\subsection{Problem Specification}
We first introduce the \textit{Maximum Quasi-clique Problem (MQCP)}.
\begin{problem}
Given a graph $G(V,E)$ and a density threshold $\alpha \in (0,1]$, find the $\alpha$-quasi-clique with the most vertices.
\end{problem}

In our paper, we study the dynamic setting. That is to say, with a fixed vertex set $V(G)$, $E(G)$ is changing by edge insertions and deletions. Specifically, we give the definition of the \textit{Dynamic Maximum Quasi-clique Problem (DMQCP)} as follows:
\begin{problem}
    Given a graph $G(V,E)$ and a density threshold $\alpha \in (0,1]$, there are two types of operations: 
    \begin{enumerate}
        \item $\texttt{AddEdge}(u,v)$: Inserting an edge $(u,v)$ to $E(G)$;
        \item $\texttt{DeleteEdge}(u,v)$: Deleting an edge $(u,v)$ in $E(G)$;
   \end{enumerate}
    After each operation, find the $\alpha$-quasi-clique with the most vertices.
\end{problem}

MQCP is known to be an NP-hard problem~\cite{pattillo2013maximum}. Therefore, its dynamic variant DMQCP is an NP-hard problem as well. In this paper, we give algorithms that can find approximate maximum quasi-cliques in near-linear time after each operation.

\subsection{Estimate Jaccard Similarity by Hashing}
\label{prelim:minhash}
The Jaccard similarity is defined as $Jaccard(A,B) = \frac{|A \cap B|}{|A \cup B|}$ for two sets $A,B$. The similarity between two vertices $u$ and $v$ is defined as $\sigma(u, v) = \frac{|N(u) \cap N(v)|}{|N(u) \cup N(v)|}$.

There are many ways to approximate the Jaccard similarity using the hash method. In our paper, we utilize two of these methods that can be adapted to the dynamic setting, subject to element insertion and deletion. Specifically, we use the $l$-Buffered-$k$-MinHash Method~\cite{10.1145/3701551.3703491} and the Bottom-$k$-MinHash Method~\cite{thorup2013bottom}. The details are as follows:
\begin{enumerate}[leftmargin=*]
    \item $l$-Buffered-$k$-MinHash Method: We generate $k$ hash values via independent hash functions for each vertex. Then, we compute the minimum hash value from $N(u)$, denoted as $\mathcal{S}_{u,i}$ with respect to the $i$-th hash function. Then we have the following formula to estimate the Jaccard similarity:
\begin{equation}
\label{equation:l_buffered_k_MinHash}
\hat{\sigma}(u,v) = \frac{\sum_{i=1}^k [\mathcal{S}_{u,i} = \mathcal{S}_{v,i}]}{k}
\end{equation}

\item Bottom-$k$-MinHash Method: We generate a distinct hash value for each vertex. Let $S_k(A)$ denote the first $k$-smallest hash values for set $A$. Then we have the following formula to estimate the Jaccard similarity:
\begin{equation}
\label{equation:bottom_k_MinHash}
\hat{\sigma}(u,v) = \frac{|S_k(N(u) \cup N(v)) \cap S_k(N(u)) \cap S_k(N(v))|}{k} 
\end{equation}
\end{enumerate}

These methods can support updating in terms of adding or deleting one element. Specifically, we give the following two lemmas summarizing their time complexity for initialization, updating, and querying $\hat{\sigma}(u,v)$ for any $u,v$.
\begin{lemma}
\label{lemma:buffer}
The time complexity of \texttt{Init}, \texttt{Update}, \texttt{Query} operation on a set $A$ in $l$-buffered-$k$-MinHash Method is $O(|A|\cdot k\log\log|A|)$, $O(k \log|A|)$, $O(k)$, respectively. The space complexity of $l$-buffered $k$-MinHash Method is $O(k \log|A|)$.
\end{lemma}

\begin{lemma}
\label{lemma:bottom}
The time complexity of \texttt{Init}, \texttt{Update}, \texttt{Query} operation on a set $A$ in Bottom $k$-MinHash-Method is $O(k+|A|\log|A|)$, $O(k +\log|A|)$, $O(k)$, respectively. The space complexity of Bottom-$k$-MinHash Method is $O(k + |A|)$.
\end{lemma}

\newcommand{\kw}[1]{{\ensuremath{\texttt{#1}}}}

\section{SIMILARITY-BASED ALGORITHM}
\label{sec:algo}
In this section, we propose a new algorithm named \texttt{DMI} that solves the DMQCP problem. In \Cref{sec:algoreview}, we first review an existing algorithm on the static graph. Then in \Cref{sec:algooverview}, we give an overview of how our algorithm works and discuss details on the initialization. In \Cref{sec:algoinsert} and \Cref{sec:algodelete}, we discuss details on insertions and deletions. Finally, we summarize the algorithm and give more insights in \Cref{sec:algoconclude}.
\subsection{Similarity-based Approach for Static Setting}
\label{sec:algoreview}
We start by reviewing the quasi-clique detection framework proposed by Pang et al.~\cite{pang2024similarity}. A critical process is to extract quasi-cliques from the neighborhood of a single vertex $u$. Since vertices within the quasi-clique tend to be more similar to one another than to vertices outside, the idea here is to select vertices with high similarities with respect to $u$. Instead of using the Jaccard similarity, Pang et al. propose to use the containment score, defined as follows:  

\begin{definition}[Containment score]
\label{def:Containment score)}
Given two vertices $u$ and $v$, the containment score of $u$ in $v$ is defined as
\begin{equation}
t(u, v) = \frac{|N(u) \cap N(v)|}{|N(u)|}
\end{equation}
\end{definition}

A straight-forward algorithm is as follows: For two parameter $\gamma$ and $b$, we select every vertex $v \in N(u)$ such that $t(u,v) \geq \gamma$ and form a candidate $C$ if $(|C|-1)/|N(u)|\geq b$, where $b$ is a fraction threshold. With exact containment scores, $\delta(C)\geq1-(1-\gamma)/b$. For $\gamma=0.9$ and $b=0.6$, the bound is $0.8333<\alpha=0.9$. \texttt{AddClique} checks each candidate's exact density against $\alpha$. 

To find large quasi-cliques, we can try extracting from every neighborhood. Pang et al. propose a vertex ordering strategy to prune unpromising candidates, which is done by computing an upper bound of quasi-cliques extractable from each neighborhood, namely the $\gamma$-degree. In turn, we try each neighborhood by the descending order of $d_\gamma(u)$ and halt when $d_\gamma(u)$ is smaller than the current best candidate.

\begin{definition}[$\gamma$-degree]
\label{def:gamma-degree}
Given a graph G and a vertex $u$, we define the $\gamma$-degree of $u$ as the number of vertices $v\in N(u)$ with $|N(v)|\geq\gamma|N(u)|$, denoted as $d_\gamma(u)$. 
\begin{equation}
d_\gamma(u) = |\{v \in N(u) \mid |N(v)| \ge \gamma \cdot |N(u)|\}|
\end{equation}
\end{definition}

\begin{algorithm}[t]
\small
\caption{\texttt{FastNBSim: Detect}}\label{alg:Detect}
\KwIn{$G$: a graph; $u$: a vertex; $\gamma, b$: two parameters $\in (0,1]$.}
\KwOut{$C$: a vertex set extracted from $N(u)$.}
\SetKwProg{Fn}{Procedure}{}{}
\SetNoFillComment

$C \leftarrow \emptyset$; \\ \label{Detect:C_init}
\If{$u$'s signature is not computed}{
    Compute the signature of $u$; \\ \label{Detect:CalcSignatureU}
}
\For{$v \in N(u)$}{
    \If{$v$'s signature is not computed}{
        Compute the signature of $v$; \\ \label{Detect:CalcSignatureV}
    }
    Calculate $\hat{\sigma}(u,v)$ via \Cref{equation:l_buffered_k_MinHash} or \Cref{equation:bottom_k_MinHash}; \\ \label{Detect:EstimateSimilarity}
    $\hat{t}(u,v) = \frac{(|N(u)|+|N(v)|)\hat{\sigma}(u,v)}{|N(u)|(1+\hat{\sigma}(u,v))}$; \\ \label{Detect:EstimateContainmentScore}
    \lIf{$\hat{t}(u,v) \geq \gamma$}{
        $C \leftarrow C \cup \{v\}$ \label{Detect:AddVertex}
    }
}
\lIf{$\frac{|C|-1}{|N(u)|} < b$}{
    $C \leftarrow \emptyset$  \label{Detect:Invalid}
}
\Return{$C$}; \\ \label{Detect:ReturnC}

\end{algorithm}

As computing the containment score could be slow, we can instead compute an approximate one. It is done by computing an estimate of Jaccard similarity $\hat{\sigma}(u,v)$. Then we convert the Jaccard similarity to the containment score $\hat{t}(u,v)$ to apply the vertex selection (\Cref{Detect:EstimateContainmentScore}). To sum up, we give the \texttt{Detect} subroutine in \Cref{alg:Detect}, which is a direct application of the aforementioned vertex selection framework(\texttt{FastNBSim}).

\subsection{Overview and Initialization}
\label{sec:algooverview}
To adapt to the dynamic setting, instead of preserving one large quasi-clique, we preserve a list of quasi-cliques $L$. In this way, for an operation of inserting or deleting, retaining alternatives can reduce dependence on any one candidate. On the other hand, as each of our quasi-clique is extracted by exploring some vertex $u$'s neighborhood $N(u)$, an edge $(u,v)$ changes the endpoint neighborhoods and can also change containment scores used by other centers. However, indirectly speaking, although a single edge change may be insignificant, a series of change could force us to update $L$:
\begin{itemize} [leftmargin=*]
    \item Some quasi-cliques from $L$ may be deleted, as too many edges are deleted to make them illegal;
    \item Some quasi-cliques may be removed from $L$, as higher quality quasi-cliques are exposed to make them insignificant;
    \item New quasi-cliques may be exposed and added to $L$, as edges are inserted to satisfy the density condition;
    \item Existed quasi-cliques may be added to $L$, as $L$ may be compromised after graph structure changes a lot.
\end{itemize}

We design a framework that is tailored to deal with these scenarios efficiently while maintaining the quality of extracted quasi-cliques, via techniques including the dynamic $k$-MinHash and batch reconstruction, called \textit{\underline{D}ynamic \underline{M}inHash-based \underline{I}ndex (\texttt{DMI})}.

\begin{algorithm}[t]
\small
\caption{\texttt{DMI: Build}}\label{alg:Build}
\KwIn{$G$: a graph; $\gamma, b, \alpha, r_{tol}$: four parameters $\in (0,1]$; $B$: capacity of candidate list $L$.}
\KwOut{$L$: a list of candidate $\alpha$-quasi-cliques.}
\SetKwProg{Fn}{Procedure}{}{}
\SetNoFillComment

$L \leftarrow \emptyset$; \\ \label{Build:L_init}
Initialize signatures and all $d_\gamma$ from $G$; reset the update counter; \\
\For{$u$ in descending $d_{\gamma}(u)$ order}{ \label{Build:TryVertex} 
    \lIf{$L\neq\emptyset$ and $d_{\gamma}(u) < \min_{S \in L}{|S|}$}{
        \textbf{break}  \label{Build:Break} 
    }
    $C \leftarrow $ \kw{Detect}$(G, u,\gamma,b)$; \\ \label{Build:Detect} 
    \kw{AddClique}$(L,C,\alpha,r_{tol},B)$; \\ \label{Build:Update} 
}

\Return $L$; \label{Build:return}

\Fn{\kw{AddClique}$(L,C,\alpha,r_{tol},B)$}{
    \lIf{$|C|<2$ or $\delta(C) < \alpha$}{
        \Return \label{AddClique:LowDensity}
    }
    $S_{r} \leftarrow C,\ S_{min} \leftarrow C$; \\ \label{AddClique:init}
    \For{$S \in L$}{ \label{AddClique:for}
        \lIf{\kw{Jaccard}$(S,C) > r_{tol}$ and $|S| < |S_{r}|$}{
            $S_{r} \leftarrow S$ \label{AddClique:S_r_update}
        }
        \lIf{$|S| < |S_{min}|$}{
            $S_{min} \leftarrow S$ \label{AddClique:S_min_update}
        }
    }
    \lIf{$|C| > |S_{r}|$}{
        $L \leftarrow L \setminus \{S_{r}\} \cup \{C\}$ \label{AddClique:replace_S_r}
    }
    \lElseIf{$|L| < B$}{
        $L \leftarrow L \cup \{C\}$ \label{AddClique:add_C}
    }
    \lElseIf{$|C| > |S_{min}|$}{
        $L \leftarrow L \setminus \{S_{min}\} \cup \{C\}$ \label{AddClique:replace_S_min}
    }
}
\end{algorithm}

In \Cref{alg:Build}, we give details on how to initialize the quasi-clique candidate list $L$ from the beginning, in the subroutine $\texttt{Build}$. In the meantime, an important subroutine $\texttt{AddClique}$ is also introduced here, which adds a newly extracted quasi-clique to $L$ following a series of rules.

\paragraph{List Construction} By calling \texttt{Build} with four parameters $\gamma, b, \alpha$ and $r_{tol}$, we build a list $L$ of a number of $\alpha$-quasi-cliques candidates. Here, $\gamma$ and $b$ guide candidate generation, $\alpha$ is enforced by exact density validation, and $r_{tol}$ guides redundancy reduction in \texttt{AddClique}. $B$ limit the total number of quasi-cliques stored in $L$. We follow the $\gamma$-degree principle too, while keeping up to $B$ candidate $\alpha$-quasi-cliques instead of the largest one. We start by intializing $L$ to $\emptyset$ in the beginning (\Cref{Build:L_init}). Then we enumerate each vertex $u$ in the descending order of $d_{\gamma}(u)$ (\Cref{Build:TryVertex}). For each vertex, if $d_{\gamma}(u)$ is smaller than the size of the current smallest quasi-clique in $L$, we halt the enumeration process, and no more quasi-cliques will be added to $L$ (\Cref{Build:Break}). Otherwise, we call $\texttt{Detect}(G,u,\gamma,b)$ to extract the quasi-clique $C$ from $u$ and $u$'s neighborhood (\Cref{Build:Detect}). Then, we try adding $C$ to $L$ by calling $\texttt{AddClique}(L,C,\alpha,r_{tol},B)$ (\Cref{Build:Update}).

\paragraph{Quasi-clique Adding} We check whether $C$ can be added to $L$ and whether any existing quasi-clique should be replaced by calling $\texttt{AddClique}(L,C,\alpha,r_{tol},B)$. We start by verifying if $C$ is a legit $\alpha$-quasi-clique as it could still display a low density over our MinHash method (\Cref{AddClique:LowDensity}). Then we use two temporary vertex sets $S_{r}$ and $S_{min}$ to find quasi-cliques that may be replaced by $C$. Specifically, $S_{r}$ stores the smallest quasi-clique with similarity over the limit $r_{tol}$, measured by the Jaccard similarity. $S_{min}$ stores the smallest quasi-clique among all in $L$. They are all initialized to be $C$ (\Cref{AddClique:init}). We now enumerate every quasi-clique $S \in L$ (\Cref{AddClique:for}) and do the following selection process: (1) We compute the Jaccard similarity between $S$ and $C$. If it is larger than $r_{tol}$ and $|S|<|S_r|$, we assign $S_r$ to $S$ (\Cref{AddClique:S_r_update}). (2) If $S$ is smaller than $S_{min}$, we assign $S_{min}$ to $S$.

Now, we update $L$ by the following process: (1) If $C$ is larger than $S_{r}$, then we replace $S_{r}$ with $C$ (\Cref{AddClique:replace_S_r}). This is because if we allow quasi-cliques with high similarity in $L$, the list can contain redundant candidates. (2) Otherwise, if there is less than $B$ quasi-cliques in $L$, then we simply add $C$ into $L$ as there is room for it (\Cref{AddClique:add_C}). (3) Otherwise, if $C$ is larger than the smallest quasi-clique $S_{min}$, we replace $S_{min}$ with $C$ (\Cref{AddClique:replace_S_min}).

In this way, we store at most $B$ distinct $\alpha$-quasi-cliques with good quality.

\paragraph{Time Complexity} The \texttt{Build} process enumerate each vertex $u$ at most once and call \texttt{Detect} on $u$ at most once. For each \texttt{Detect} process, its complexity is dominated by enumerating the neighborhood. The total cost for all \texttt{Detect} is $O(n+m)$. As each vertex's signature is also computed exactly once, by \Cref{lemma:buffer} and \Cref{lemma:bottom}, the total time complexity for all \texttt{Detect} is $\tilde{O}(n+m)$\footnote{$\tilde{O}$ hides a factor of $O(poly(\log n))$} as we treat $k$ as a constant. We only aim at keeping a constant number of candidate quasi-cliques in $L$. Exact density validation and candidate comparisons must also be counted. Let $\tau(C)$ be the cost of \texttt{AddClique} for a tested candidate $C$. Then $T_{\mathrm{Build}}=\tilde{O}(n+m)+\sum_C\tau(C)$.

\subsection{Insertion}
\label{sec:algoinsert}
In this section, we discuss how to update the candidate list $L$ with an edge insertion of $(u,v)$. Since now we have the $\texttt{AddClique}$ subroutine to determine if the newly extracted quasi-clique can be added into $L$, for any insertion, we only need to find the proper quasi-clique to add in, addressing the two scenarios aforementioned at the beginning of \Cref{sec:algooverview}.

\begin{algorithm}[t]
\small
\caption{\texttt{DMI: AddEdge}}\label{alg:AddEdge}
\KwIn{$G$: a graph; $(u,v)$: the edge to add; $L$: a list of candidate $\alpha$-quasi-cliques; $\gamma, b, \alpha, r_{tol}$: four parameters $\in (0,1]$; $B$: capacity of candidate list $L$; $Batch$ : the number of operations to trigger rebuild.}
\KwOut{$L'$: an updated list of candidate $\alpha$-quasi-cliques.}
\SetKwProg{Fn}{Procedure}{}{}
\SetNoFillComment

$E(G) \leftarrow E(G) \cup \{(u,v)\}$; Increment the update counter; \\ \label{AddEdge:E(G)_update}
\If{$\#$ of operations $=Batch$}{ \label{AddEdge:Rebuild_0}
    $L \leftarrow $ \kw{Build} $(G,\gamma,b,\alpha,r_{tol},B)$; \\ \label{AddEdge:Rebuild_1}
    \Return $L$; \label{AddEdge:return_rebuild}
}

% $L' \leftarrow \emptyset$; \\ \label{AddEdge:L'_init}

Add $v$ to signature of $u$; \\ \label{AddEdge:u_kminhash_update}
Add $u$ to signature of $v$; \\ \label{AddEdge:v_kminhash_update}
Recompute $d_\gamma(u)$ and $d_\gamma(v)$ by \Cref{def:gamma-degree}; \\ \label{AddEdge:endpoint_gamma}

\For{$x \in \{u,v\}$}{
    \For{$w \in N(x)\setminus\{u,v\}$}{
        \If{$|N(x)|-1<\gamma|N(w)|\leq|N(x)|$}{ \label{AddEdge:ifgamma}
            $d_{\gamma}(w) \leftarrow d_{\gamma}(w) + 1$; \\ \label{AddEdge:d_gamma_w_update}
            \If{$L=\emptyset$ or $d_{\gamma}(w) \geq \min_{S \in L}{|S|}$}{ \label{AddEdge:ifbetter}
                $C \leftarrow $ \kw{Detect}$(G,w,\gamma,b)$; \\ \label{AddEdge:Detect} 
                \kw{AddClique}$(L,C,\alpha,r_{tol},B)$; \\ \label{AddEdge:Update} 
            }
        }
    }
}

{
\For{$x\in\{u,v\}$}{
    \If{$L=\emptyset$ or $d_\gamma(x)\geq\min_{S\in L}|S|$}{
        $C\leftarrow\kw{Detect}(G,x,\gamma,b)$; \\
        $\kw{AddClique}(L,C,\alpha,r_{tol},B)$; \\
    }
}
}
\lFor{$S \in L$ with $u,v\in S$}{
    $E(S) \leftarrow E(S) \cup \{(u,v)\}$ \label{AddEdge:AddEdgeToClique} 
}

% $L' \leftarrow L$; \\ \label{AddEdge:NewSolutionSet} 

\Return $L$; \label{AddEdge:return}
\end{algorithm}

We provide the implementation details of $\texttt{AddEdge}(u,v)$ in \Cref{alg:AddEdge}. Note that parameters including $\gamma, b, \alpha, r_{tol}, B$ and $Batch$ are carried along among all subroutines. We omit discussing them again here for brevity. In the general initialization (\texttt{Build}) process, we have computed the signature of each vertex, as well as their $\gamma$-degree. After inserting an edge, we have to update them correspondingly. Furthermore, an important aspect in our algorithm is the reconstruction process. To avoid being biased, after $Batch$ operations (including both insertions and deletions), we will rebuild the whole candidate list by calling \texttt{Build} again (\Cref{AddEdge:Rebuild_0} to \Cref{AddEdge:return_rebuild}). Triggering rebuilding too often could cause slower runtime in general, while the opposite may compromise the quasi-clique quality. This time-quality trade-off will be evaluated empirically in the experiment section. 

% To begin with, we initialize $L'$ to be $\emptyset$ in the beginning (\Cref{AddEdge:L'_init}. The updated $L$ will be stored in it and returned as the result in the end. 
We add $(u,v)$ into $E(G)$ (\Cref{AddEdge:E(G)_update}) at first. The edge is applied before any scheduled rebuild, so that rebuilding uses the updated graph. Since $(u,v)$ is added, we need to add $u$ to $v$'s signature as well as adding $v$ to $u$'s signature (\Cref{AddEdge:u_kminhash_update}-\Cref{AddEdge:v_kminhash_update}). This can be done efficiently by the dynamic $k$-MinHash technique elaborated in \Cref{prelim:minhash}. We recompute the endpoints' $\gamma$-degrees directly, since their own thresholds change. We then update other neighbors incrementally, excluding $u,v$ to avoid double-counting, and also test the endpoints for extraction. For each vertex $w \in N(x)\setminus\{u,v\}$, $x \in \{u,v\}$, if $w$'s $\gamma$-degree is increased due the edge insertion (\Cref{AddEdge:ifgamma}), we increase $w$'s $\gamma$-degree by $1$ (\Cref{AddEdge:d_gamma_w_update}). Then we check if $L$ is empty or $d_{\gamma}(w)$ is no less than the size of the smallest quasi-clique in $L$ (\Cref{AddEdge:ifbetter}), indicating there is a potential of finding a better one. If so, we try extracting the quasi-clique from $w$'s neighborhood (\Cref{AddEdge:Detect}) and update $L$ with it (\Cref{AddEdge:Update}). For each quasi-clique $S \in L$, we add $(u,v)$ into $E(S)$ if necessary (\Cref{AddEdge:AddEdgeToClique} ). In the end, we return $L$ as the result (\Cref{AddEdge:return}). Note that for the query itself, we simply return the largest $\alpha$-quasi-clique in $L$ (or its size) as it is naturally maintained.

\paragraph{Time Complexity} For a non-rebuild insertion, endpoint recomputation adds $O(d(u)+d(v))$ time. At most $O(d(u)+d(v))$ centers are tested. The neighborhood and signature work is $\tilde{O}(d_{max}^2)$, with an additional $\sum_C\tau(C)$ for exact validation and candidate comparisons. A scheduled reconstruction adds $T_{\mathrm{Build}}$ to the triggering update, or an average of $T_{\mathrm{Build}}/Batch$ per operation.

\subsection{Deletion}
\label{sec:algodelete}
Deletion may cause quasi-cliques to become invalid, i.e., the edge-density becomes less than $\alpha$. We either delete them directly, or check if by removing a few number of vertices, it can become $\alpha$-quasi-clique again. In worst case, we search for other quasi-cliques and update the candidate list $L$.

\begin{algorithm}[t]
\small
\caption{\texttt{DMI: DeleteEdge}}\label{alg:DeleteEdge}
\KwIn{$G$: a graph; $(u,v)$: the edge to delete; $L$: a list of candidate $\alpha$-quasi-cliques; $\gamma, b, \alpha, r_{tol}$: four parameters $\in (0,1]$; $B$: capacity of candidate list $L$; $Batch$ : the number of operations to trigger rebuild.}
\KwOut{$L'$: an updated list of candidate $\alpha$-quasi-cliques.}
\SetKwProg{Fn}{Procedure}{}{}
\SetNoFillComment

$E(G) \leftarrow E(G) \setminus \{(u,v)\}$; Increment the update counter; \\ \label{DeleteEdge:E(G)_update}
\If{$\#$ of operations $=Batch$}{ \label{DeleteEdge:Rebuild_0}
    $L \leftarrow $ \kw{Build} $(G,\gamma,b,\alpha,r_{tol},B)$; \\ \label{DeleteEdge:Rebuild_1}
    \Return $L$; \label{DeleteEdge:return_rebuild}
}

% $L' \leftarrow \emptyset$; \\ \label{DeleteEdge:L'_init}

Delete $v$ to signature of $u$; \\ \label{DeleteEdge:u_kminhash_update}
Delete $u$ to signature of $v$; \\ \label{DeleteEdge:v_kminhash_update}
Recompute $d_\gamma(u)$ and $d_\gamma(v)$ by \Cref{def:gamma-degree}; \\ \label{DeleteEdge:endpoint_gamma}

\For{$x \in \{u,v\}$}{ \label{DeleteEdge:for}
    \For{$w \in N(x)\setminus\{u,v\}$}{
        \If{$|N(x)| < \gamma \cdot |N(w)| \leq |N(x)|+1$}{
            $d_{\gamma}(w) \leftarrow d_{\gamma}(w) - 1$; \\ \label{DeleteEdge:d_gamma_w_update}
        }
    }
}

\For{$S \in L$}{
    \lIf{\kw{CliqueDeleteEdge}$(S,(u,v), \alpha) = 3$}{
        $L \leftarrow L \setminus \{S\}$ \label{DeleteEdge:DeleteEdgeToClique} 
    }
}

\lIf{$L = \emptyset$}{
    $L \leftarrow $ \kw{Build} $(G,\gamma,b,\alpha,r_{tol},B)$ \label{DeleteEdge:Rebuild_2}
}

% $L' \leftarrow L$; \\ \label{DeleteEdge:NewSolutionSet} 

\Return $L$; \label{DeleteEdge:return}
\end{algorithm}

The implementation details of $\texttt{DeleteEdge}(u,v)$ are given in \Cref{alg:DeleteEdge}. The same as in \texttt{AddEdge}, to avoid being biased, we will rebuild the entire candidate list by calling \texttt{Build} again after $Batch$ operations (\Cref{DeleteEdge:Rebuild_0}-\Cref{DeleteEdge:return_rebuild}). We first delete $(u,v)$ from $E(G)$ (\Cref{DeleteEdge:E(G)_update}). Then, we delete $u$ in $v$'s signature and delete $v$ in $u$'s signature (\Cref{DeleteEdge:u_kminhash_update}-\Cref{DeleteEdge:v_kminhash_update}). We recompute both endpoints' $\gamma$-degrees from the definition. By enumerating each neighbor $w$ from $N(x)\setminus\{u,v\}$, $x \in \{u,v\}$, we update their $\gamma$-degree (\Cref{DeleteEdge:for}-\Cref{DeleteEdge:d_gamma_w_update}): When $\gamma\cdot |N(w)|$ lies in $(|N(x)|,|N(x)|+1]$, decrease $d_\gamma(w)$ by $1$.

The most critical part in \texttt{DeleteEdge} is to delete the edge in each quasi-clique (\Cref{DeleteEdge:DeleteEdgeToClique}). We design \texttt{CliqueDeleteEdge} to update each quasi-clique $S$ accordingly and report the number of vertices actually removed, or a failure sentinel. Specifically, if \texttt{CliqueDeleteEdge} returns $3$, meaning that the tested local repairs failed, we remove $S$ from $L$ (\Cref{DeleteEdge:DeleteEdgeToClique}). In addition, if $L$ becomes empty after deletion, we rebuild it as well (\Cref{DeleteEdge:Rebuild_2}). In the end, we return the updated $L$ (\Cref{DeleteEdge:return}).

\begin{algorithm}[t]
\small
\caption{\texttt{DMI: CliqueDeleteEdge}}\label{alg:CliqueDeleteEdge}
\KwIn{$S$: a quasi-clique; $(u,v)$: the edge to delete; $\alpha$: a parameter $\in (0,1]$.}
\KwOut{$num$: $0$, $1$, or $2$ vertices removed, or $3$ for failed local repair.}
\SetKwProg{Fn}{Procedure}{}{}
\SetNoFillComment

\lIf{$(u,v) \notin E(S)$}{
    \Return{$0$} \label{CliqueDeleteEdge:return0_1}
}

$E(S) \leftarrow E(S) \setminus \{(u,v)\}$; \\ \label{CliqueDeleteEdge:delete_edge}

\lIf{$\delta(S) \geq \alpha$}{
    \Return{$0$} \label{CliqueDeleteEdge:return0_2}
}

$x \leftarrow u,\ y \leftarrow v$; \\ \label{CliqueDeleteEdge:init_xy}
\lIf{$|N(x,S)| > |N(y,S)|$}{
    \kw{swap}$(x,y)$ \label{CliqueDeleteEdge:swap_xy}
}

\For{$w \in S\setminus\{u,v\}$}{
    \lIf{$|N(w,S)| < |N(x,S)|$}{
        $y \leftarrow x,\ x \leftarrow w$ \label{CliqueDeleteEdge:update_mindeg_1}
    }
    \lElseIf{$|N(w,S)| < |N(y,S)|$}{
        $y \leftarrow w$ \label{CliqueDeleteEdge:update_mindeg_2}
    }
}

\For{$w \in \{u,v,x\}$}{
    \lIf{$|S\setminus\{w\}|\geq2$ and $\delta(S \setminus \{w\}) \geq \alpha$}{
        $S \leftarrow S \setminus \{w\}$, 
        \Return $1$\label{CliqueDeleteEdge:erase_one}
    }
}

\For{distinct pairs $\{z,w\} \subseteq \{u,v,x,y\}$}{
    \lIf{$|S\setminus\{z,w\}|\geq2$ and $\delta(S \setminus \{z,w\}) \geq \alpha$}{
        $S \leftarrow S \setminus \{z,w\}$, 
        \Return $2$ \label{CliqueDeleteEdge:erase_two}
    }
}

\Return $3$; \label{CliqueDeleteEdge:return3}
\end{algorithm}

Now we discuss details of \texttt{CliqueDeleteEdge}. The implementation details is given in \Cref{alg:CliqueDeleteEdge}. If $(u,v) \notin E(S)$, no vertex needs to be deleted and thus return $0$ (\Cref{CliqueDeleteEdge:return0_1}). Otherwise, we remove $(u,v)$ from $E(S)$ (\Cref{CliqueDeleteEdge:delete_edge}). Then, we check if $\delta(S)$ is still at least $\alpha$, indicating its validity. If so, we return with $0$ vertex that needs to be removed (\Cref{CliqueDeleteEdge:return0_2}). In the following, we check if removing one or two vertices can guarantee the edge density. We first extract the two distinct vertices with the minimum number of neighbors, stored in $x$ and $y$ (\Cref{CliqueDeleteEdge:init_xy}-\Cref{CliqueDeleteEdge:update_mindeg_2}). We first check if removing one vertex is enough. In this case, we remove either a vertex from $u,v$, or the one with smallest neighborhood size ($x$). We try each of them and see if the density condition can be satisfied. If so, we remove this vertex and return $1$ (\Cref{CliqueDeleteEdge:erase_one}). Then, we check if removing two vertices is enough; similarly, we try every pair from $\{u,v,x,y\}$ (vertices from $(u,v)$ and vertices with the smallest neighborhood size). If the density can be satisfied, we remove them and report with $2$ (\Cref{CliqueDeleteEdge:erase_two}). For any other case, we report with $3$ to indicate failure of the tested local repairs (\Cref{CliqueDeleteEdge:return3}).

Below, we give sufficient conditions under which removing one vertex restores the density after an edge deletion. The proof follows by degree-counting arguments and is omitted due to space.

% lemma 用于说明在大多数情况下都只需要删除 1 个点（或者2个点），如果要删除3个点以上，那么将认为这次删除是伤筋动骨的，需要彻底从备选列表里清除。
% 假如(1)和(2)均不满足，d_S(a_1)的取值范围最多只有一个正整数。
\begin{lemma}
\label{lemma:delete}
Let $S$ be an $\alpha$-quasi-clique with $s\geq3$, and $a_1,a_2,\dots,a_s$ are vertices of $S$ in the non-decreasing order of $d_S(a_i)$. All degrees and edge counts are measured before deleting $(u,v)$. After deleting edge $(u,v)$ from $S$, there exists $S' = S \setminus \{w\}$ that $S'$ is an $\alpha$-quasi-clique when satisfying one of the following constraints: (1) $d_S(a_1) \geq \alpha(s-1)$; (2) $d_S(a_1) \leq \frac{\sum_{i=1}^s d_S(a_i)}{s} - 1$; (3) $d_S(u)=d_S(a_1)$ or $d_S(v) = d_S(a_1)$.
\end{lemma}

% 在删除两个点的情况，我们只需要考虑 u,v 和两个度数最小的点中挑两个进行删除，因为这样删除的总边数最小。
(1) means that the minimum degree is at least the required average degree, indicating that all vertices have a large degree; (2) means that the minimum degree is at most the current average degree minus $1$, ruling out the case where every vertex has the same degree; (3) refers to the case when $u$ or $v$ is the vertex with the minimum degree; In any of these cases, we can remove one vertex to satisfy the edge-density condition. In addition, since enumerating the case that removes two vertices is acceptable in terms of efficiency, we add this logic in \texttt{CliqueDeleteEdge} to enhance the robustness. If none of the tested removals succeeds, we discard this candidate.

% 证明这个 lemma 的意义是说明大多数情况下最多只用删除一个点
% case1 是最小的deg大于等于需要的平均度数，大家的deg都很大就只用删除 x 
% case2 是最小的deg 不大于 平均deg减1，只要度数分配稍微不平均一点就可以仅仅删除最小点
% case3 是假如 u 或者 v 是度数最小的点之一，删去它就可以

% 探测到删除两个点以内的情况是因为时间复杂度恰好够（恰好cover住），如果需要删除3个点以上，我们就认为它的结构已经遭到了“伤筋动骨”的破坏，无法收缩成\delta(S)>=alpha的情况，需要从List中剔除

\paragraph{Time Complexity} For a non-rebuild deletion, recomputing both endpoint scores and updating affected neighbors costs $O(d(u)+d(v))$. With maintained induced-edge counts and vertex degrees, finding the two minimum-degree vertices and testing the constant number of repairs costs $O(|S|)$ per candidate. Since $B$ and $\alpha$ are fixed and feasible candidates have $|S|=O(d_{max})$, the non-rebuild cost is $\tilde{O}(d_{max})$. Each scheduled or empty-list reconstruction adds $T_{\mathrm{Build}}$.

\subsection{Overall Analysis}
\label{sec:algoconclude}
From \Cref{sec:algooverview}, \Cref{sec:algoinsert} and \Cref{sec:algodelete}, the total runtime includes neighborhood processing, exact candidate validation and every reconstruction.
% However, in practice, it is convincing to give a much faster performance, for the following reasons:
% \begin{itemize}[leftmargin=*] 
%     \item One edge insertion would not likely cause a large number of vertices to have a good potential of being the center of large quasi-cliques since the $\gamma$-degree only increases at most two each time, and we only call \texttt{Detect} and \texttt{AddClique} when it exceeds the size of the smallest quasi-clique in $L$. Therefore, we only call the extraction process in very few scenarios. 
%     \item One edge deletion would not likely cause a massive rebuild or quasi-cliques disappearing, since the $\gamma$-degree only decreases at most two each time, and we prove that in most scenarios only one vertex needs to be deleted to guarantee the edge-density.
%     \item In real-world data, only a small number of vertices have very large degrees.
%     \item By setting $batch$ to an appropriately large number, the rebuild cost can be amortized into each operation. We also choose to keep the candidate list $L$'s size remains constant, i.e., $B$ is $O(1)$.
% \end{itemize}
Since we only keep $O(1)$ quasi-cliques in $L$, and the $k$ parameter in the dynamic MinHash method is also constant, the memory is $\tilde{O}(n+m)$.

\section{NEIGHBOR-SEARCH-BASED ALGORITHM}
\label{sec:neighbor}
In this section, we propose a side-product method, \texttt{NSF}, which directly uses the neighbor detection routine \texttt{Detect} of Pang et al.~\cite{pang2024similarity} during the update. While \texttt{DMI} outperforms it in the empirical evaluation, it is applicable compared to other baselines. 

The idea is that for any edge update of $(u,v)$, it only affects the neighborhood of $u$ and $v$. Since \texttt{Detect} is mainly based on searching over the neighborhood of a center. If we keep track of the quasi-clique extracted from every vertex's neighborhood, we only need to re-extract from all affected vertices, i.e., $u$ and $v$'s neighbors. Such a method is an improved version of calling the whole extraction algorithm again whenever an update happens. However, it does not fully address the change caused by the edge update. For example, inserting an edge could enhance some other quasi-cliques which do not exhibit a density of $\geq \alpha$ originally. For efficiency, this method does not account for such a scenario. \texttt{NSF} stores a candidate for each vertex, while \texttt{DMI} maintains MinHash signatures and a bounded candidate list. Their peak memory usage is compared in \Cref{tab:rand_memoryuse_results,tab:NSF_rand_memoryuse_results}.

For initialization, we call the \texttt{Detect} routine for every vertex $u$, storing the extracted quasi-clique at $C(u)$. If the found quasi-clique's density is at least $\alpha$, we store its information into an ordered set $S$ for looking up. For any update, we insert or delete the edge $(u,v)$ from the graph, as well as erasing the information of $C(u)$ and $C(v)$ from $S$. After calling \texttt{Detect} on $u$ and $v$ again, we re-insert their information into $S$. Each time, \texttt{Extract} re-detects and validates candidates in descending stored-size order, using at most $R$ attempts, and returns the best feasible one encountered. The implementation details are included in the artifact repository.

% build的流程就是 for 每个节点，然后跑对应的邻域探测算法，C[u]数组存放的是u这个节点的邻域找到的准团（伪代码这里写的是节点到准团的映射），如果对应的准团密度大于等于alpha就将（准团大小，节点u）这个二元组放进 set 中，set S采用平衡树维护。

\section{EVALUATION}
\label{sec:experiment}
In this section, we provide a comprehensive empirical evaluation on the efficiency and solution quality of our proposed \texttt{DMI}, as well as statistics on the neighbor-search-based framework \texttt{NSF}. \Cref{sec:furtherexperiment} reports seven experiments on hyperparameters, operation-size scalability, stage-wise time cost, candidate-list ablation, memory usage, rebuild overhead and update latency, and discard/rebuild statistics.
% Specifically, we address the following research questions to evaluate various important aspects of our algorithms:
%  \begin{itemize}[leftmargin=*]
%      \item \textbf{RQ1 (Efficiency):} Given graphs with both insertions and deletions, what are the runtimes for our methods and the baselines, and how do they scale with very large graphs? 
%      \item \textbf{RQ2 (Efficiency on Special Case):} Given graphs with edge updates following a certain pattern, i.e., only insertions/deletions, temporal graphs, what are the runtimes for our methods and the baselines, and is there any difference compared to the general case?
%      \item \textbf{RQ3 (Effectiveness):} What are our methods’ and baselines’ performances in finding the $\alpha$-quasi-cliques over all updates?
%      \item \textbf{RQ4 (Hashing Techniques):} What are the comparison result between using the $l$-buffered $k$-MinHash and Bottom $k$-MinHash and what's the insight from them?
%  \end{itemize}

% We now present experimental results. We first discuss the setup
% in \Cref{exp:setup}, then discuss the results of our methods against the baseline algorithms. We give some detailed analysis
% of the effect of parameters and pruning techniques.
% \todo{it seems manny compile error, please fix!!!!!!!!!}
\subsection{Experimental Setting}
\label{exp:setup}

\begin{table}[t]
\centering
\caption{Datasets used in experiments.}
\label{tab:datasets-table}
\begin{tabular}{|c|c|c|c|c|}
\hline
Dataset & Full Name & $n$ & $m$ & $Q$ \\
\hline
FB & Ego-facebook & 4039 & 88234 & 58822 \\
HP & Ca-HepPh & 12008 & 118489 & 78992 \\
CM & Ca-CondMat & 23133 & 93439 & 62292 \\
ER & Email-Enron & 36692 & 183831 & $10^5$ \\
SF & Web-Stanford & 281903 & 1992636 & $10^5$ \\
BS & Web-BerkStan & 685230 & 6649470 & $10^5$ \\
GG & Web-Google & 875713 & 4322051 & $10^5$ \\
PK & Soc-Pokec & 1632803 & 22301964 & $10^5$ \\
DB & dblp\_coauthor & 1824701 & 1 & $10^5$ \\
\hline
\end{tabular}
\end{table}

\paragraph{Datasets} One of the difficulties in conducting experiments is to generate the dynamic graph datas. We select nine publicly-available real-world\footnote{From konect.cc and snap.stanford.edu} datasets. These graphs come without update operations. Therefore, we need to preprocess them to generate the suitable datasets. Specifically, we consider the following few types of operation generation methodology:

\begin{itemize}[leftmargin=*]
    \item At most $10^5$ insertions and deletions are generated universally randomly. No duplicate edge is inserted, and only delete edges that currently exist.
    % , as we terminate our generation process once the graph reaches fully connected. 
    These data are marked with a "rand\_" prefix.
    \item In dynamic graphs, it is common to consider the incremental and the decremental setting, i.e., only edge insertions or deletions. We uniformly sample at most $10^5$ non-existed edges to insert one by one and mark them with "inc\_" prefix. Similarly, we uniformly sample at most $10^5$ existed edges to delete, marked with "del\_".
    \item One of the well-studied dynamic graphs is the temporal graphs, with timestamps indicating the exist time-window for each edge. We preprocess them by turning the time-window information into corresponding insertions and deletions. 
    % For example, for an edge $(u,v)$ with timestamp $[s,t]$, we add an insertion $\texttt{AddEdge}(u,v)$ at time $s$ and a deletion $\texttt{DeleteEdge}(u,v)$ at time $t$. 
    If temporal graphs lack an end-time timestamp, we generate one first. To separate them, we mark them with "temp\_" and "tinc\_" prefixes. We also limit the number of operations to $10^5$.
\end{itemize}

Non-temporal-based data are mainly generated from SNAP~\cite{leskovec2014snap} while temporal-based datas are from KONECT~\cite{kunegis2013konect}. \Cref{tab:datasets-table} shows the statistics of these data. We use $n$, $m$, and $Q$ to denote the number of vertices, the number of edges at the beginning, and the total number of operations, respectively.

\paragraph{Baselines} We compare our method with the baselines from highly
 related works, where the core ideas are as follows.

\begin{itemize}[leftmargin=*]
    \item \texttt{FastNBSim}~\cite{pang2024similarity}: \texttt{FastNBSim} is the current state-of-the-art method in finding the maximum quasi-clique on a static graph. It is a similarity-based quasi-clique detection algorithm that incrementally updates neighborhood similarities to efficiently identify dense subgraphs. 
    % It enhances the original neighborhood-similarity approach through optimized pruning and candidate filtering strategies, effectively reducing redundant similarity computations and accelerating quasi-clique discovery. For accelerating the similarity comparison, it utilizes a classical $k$-MinHash method to compute the approximate containment score pairwisely.
    \item \texttt{NBSim}~\cite{pang2024similarity}: \texttt{NBSim} serves as the baseline similarity-based quasi-clique enumeration algorithm. It computes the exact pairwise neighborhood similarities to form candidate sets within the examined neighborhoods.
    \item \texttt{NuQClq}~\cite{chen2021nuqclq}: It is a local search–based algorithm for maximum quasi-clique detection. It iteratively refines candidate vertex sets using heuristic and density-guided neighborhood updates.
    % , efficiently locating highly dense subgraphs under a given quasi-threshold parameter
    % that defines the minimum edge density required for a quasi-clique. 
    Although computationally expensive, it provides a strong reference for detection accuracy.
    \item \texttt{DMI-bf} and \texttt{DMI-bt}: Two versions of our algorithm that utilize $l$-buffered $k$-MinHash and Bottom $k$-MinHash.
    \item \texttt{NSF-fn} and \texttt{NSF-ns}: Two versions of our neighbor-search-based framework, utilizing the \texttt{Detect} routine from \texttt{FastNBSim} and \texttt{NBSim}, respectively.
\end{itemize}

\paragraph{Settings} For each edge update in the dynamic graph, (\texttt{FastNBSim}, \texttt{NBSim}, and \texttt{NuQClq}) are re-invoked from scratch. For \texttt{NuQClq}, the quasi-threshold (density parameter) is set equal to the average density of the quasi-cliques obtained by our proposed algorithm on the same dataset. For both \texttt{DMI-bf} and \texttt{DMI-bt} methods, we set the parameters as $\alpha=0.9$, $\gamma = 0.9$, $b = 0.6$, $r_{tol}=0.7$, $k = 8$, $l = 8$, $Batch = 5000$, and $B = 5$. For \texttt{NSF-fn} and \texttt{NSF-ns}, we set $Batch = 5000$, $R = 20$.
Runtime is measured as the total time for all $Q$ updates, including rebuilds. Unless otherwise specified, these parameter settings are used by default. We implement all the algorithms in C++ and run experiments on a machine having an Intel(R) Xeon(R) Platinum 8358 CPU @2.60GHz and 512GB of memory, with Ubuntu installed.

% All algorithms mentioned above are implemented in C++.

% \subsection{Overall Comparison Results}
% This experiment is designed to study the efficiency and scalability of our method (\textbf{RQ1},\textbf{RQ2}) as well as the solution quality (\textbf{RQ3}). In addition, since we conduct experiments for both \texttt{DMI-bf} and \texttt{DMI-bt}, we also get insights for using different hashing techniques (\textbf{RQ4}). We compare ours with all the three competitors (\texttt{FastNBSim}, \texttt{NBSim} and \texttt{NuQClq}). In all, the empirical results demonstrate both the effectiveness and the efficiency of our method in real-world data.

\subsection{Comparison on Random Dynamic Graphs} The first set of experiments evaluates the algorithms on datasets of uniformly random edge insertions and deletions.

\begin{table}[t]
\centering
\caption{Comparison of the quasi-clique density and size across datasets for different algorithms.}
\label{tab:rand_Quasi_clique_results_transposed}
\scriptsize
\setlength{\tabcolsep}{1pt}
\renewcommand{\arraystretch}{1.1}
\resizebox{\columnwidth}{!}{%
\begin{tabular}{l|cc|cc|cc|cc|cc|cc}
\hline
 & \multicolumn{2}{c|}{rand\_BS}
 & \multicolumn{2}{c|}{rand\_ER}
 & \multicolumn{2}{c|}{rand\_GG}
 & \multicolumn{2}{c|}{rand\_HP}
 & \multicolumn{2}{c|}{rand\_SF}
 & \multicolumn{2}{c}{rand\_FB} \\
\cline{2-13}
\textbf{Algorithm}
 & Size & Dens. & Size & Dens. & Size & Dens. & Size & Dens. & Size & Dens. & Size & Dens. \\
\hline
\texttt{DMI-bt} & 162.0 & 1.00 & 11.8 & 0.92 & 45.0 & 1.00 & 223.0 & 0.99 & 65.0 & 0.95 & 113.0 & 0.94 \\
\texttt{DMI-bf} & 190.0 & 1.00 & 12.1 & 0.93 & 48.0 & 0.99 & 129.0 & 0.99 & 64.0 & 0.92 & 91.0 & 0.93 \\
\texttt{NSF-fn} & 190.0 & 1.00 & 17.0 & 0.74 & 47.8 & 0.99 & 86.0 & 0.98 & 64.0 & 0.92 & 79.2 & 0.93 \\
\texttt{NSF-ns} & 202.0 & 1.00 & 8.9 & 0.98 & 48.0 & 0.99 & 238.6 & 0.99 & 68.0 & 0.99 & 19.4 & 0.99 \\
\texttt{FastNBSim} & 190.0 & 1.00 & 17.0 & 0.74 & 47.8 & 0.99 & 86.0 & 0.98 & 64.0 & 0.92 & 79.2 & 0.93 \\
\texttt{NBSim} & 202.0 & 1.00 & 9.5 & 0.99 & 48.0 & 0.99 & 238.6 & 0.99 & 67.0 & 0.99 & 19.4 & 0.99 \\
\texttt{NuQClq} & 89.0 & 0.99 & 27.0 & 0.94 & 36.2 & 0.99 & 241.2 & 0.98 & 72.2 & 0.95 & 133.0 & 0.94 \\
\hline
\end{tabular}%
}
\end{table}

\begin{figure}[t]
    \centering
    %\vspace{0.3cm}
    %\hspace*{-0.53cm} % 向左移动 0.5 cm，可调为 -1cm、-2cm
    \includegraphics[width=0.95\columnwidth,pagebox=cropbox]{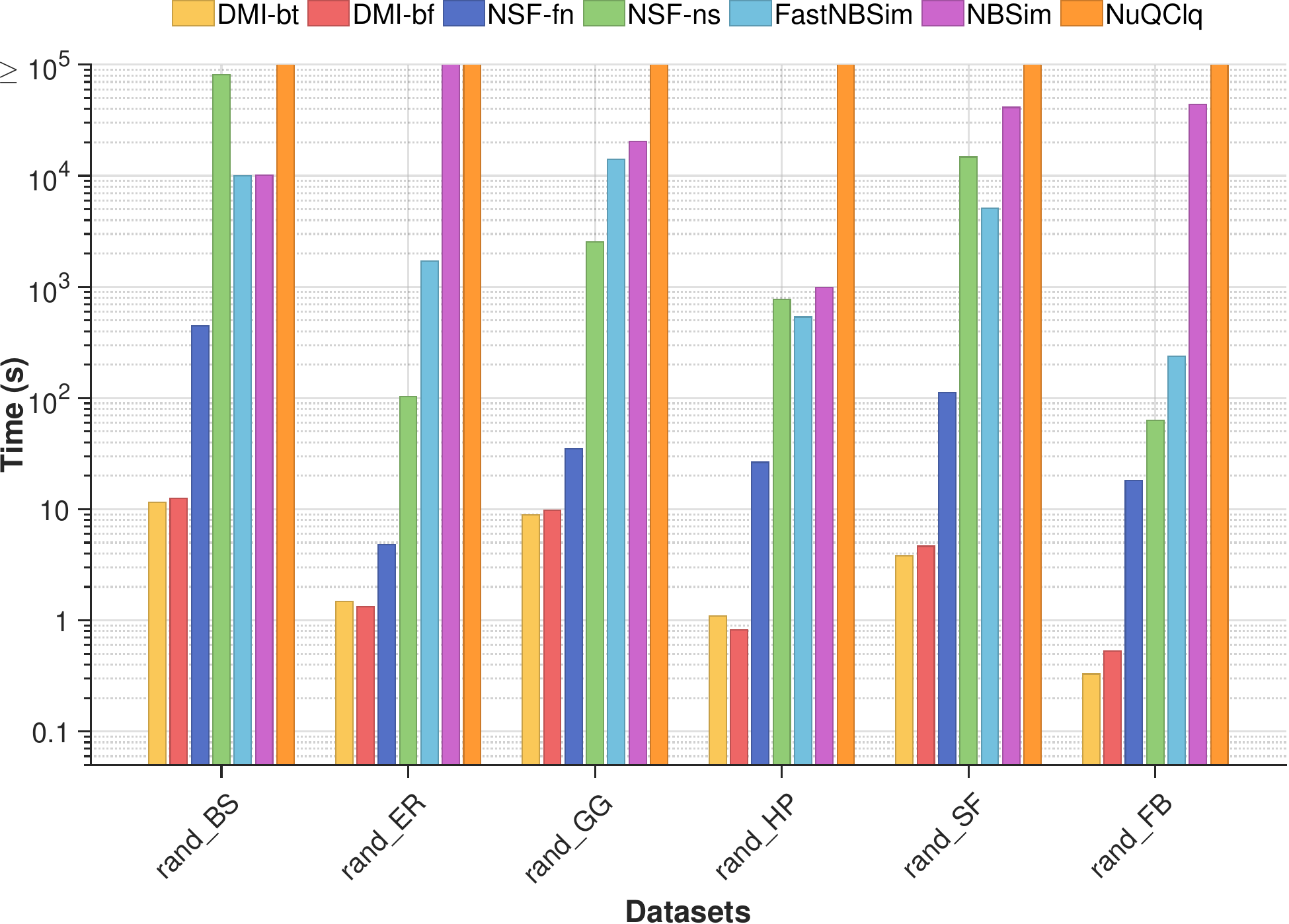} % 放大比例
    \caption{The total runtime (seconds) for all $Q$ updates for different algorithms on datasets generated from uniformly randomly edge insertions and deletions.}
    \label{fig:rand_xx time outcome}
    %\vspace{0.3cm}
\end{figure}

\subsubsection{Efficiency} We compare the total runtime of all algorithms for $Q$ operations, shown in \Cref{fig:rand_xx time outcome}. The results reveal that both \texttt{DMI-bt} and \texttt{DMI-bf} dramatically outperform all other baselines across all datasets, often by several orders of magnitude. 
\texttt{DMI-bt} and \texttt{DMI-bf} share similar performance in terms of efficiency, displaying a short total runtime, typically finishing within 10 seconds except rand\_BS (about 12--13 seconds). In contrast, \texttt{NBSim} and \texttt{FastNBSim} require hundreds to tens of thousands of seconds; \texttt{NBSim} reaches $\geq10^5$ seconds on rand\_ER. These results indicate that both variants of \texttt{DMI} are approximately $100\times$ to over $10,000\times$ faster than them. 

On the other hand, \texttt{NuQClq}'s total runtime reaches the $\geq10^5$-second level on all shown datasets. This indicates that these methods are not suitable for the dynamic scenario. For the \texttt{NSF} framework, \texttt{NSF-ns} is still very slow, since it uses the same \texttt{Detect} routine from \texttt{NBSim}. \texttt{NSF-fn} is faster but still up to two orders of magnitude slower than \texttt{DMI}.
% Similarly, on rand\_ER, rand\_PK, \texttt{NBSim} also reaches an average runtime per operation that is greater than $10^5$ seconds. These two algorithms are not scalable on the dynamic quasi-clique problem, especially on graphs with more than $10^5$ vertices and $10^6$ edges.

% While on rand\_CM, \texttt{DMI-bt} is faster than \texttt{DMI-bf}, on other random datasets, their average runtime is very close to each other. This leads to yet another key observation from statistics - the consistent run-time stability of \texttt{DMI} algorithms across all datasets. The small runtime gap indicates that both dynamic MinHash algorithms benefit equally from DMI’s efficient dynamic maintenance mechanism. We can also observe the logarithmic stability of the \texttt{DMI} methods. 

% Overall, \texttt{DMI} demonstrates up to four orders of magnitude faster than baselines, proving its efficiency and scalability. Their stable and low runtimes across diverse random dynamic graphs confirm the robustness of the \texttt{DMI} framework for handling large-scale, continuously evolving networks.

\subsubsection{Solution Quality} We compute the average density and size of the detected quasi-clique for all operations on the same dataset above, shown in \Cref{tab:rand_Quasi_clique_results_transposed}. We can see that both \texttt{DMI-bf} and \texttt{DMI-bt} consistently produce high-quality quasi-cliques of $\delta(S)\in[0.91,1.00]$, with similar size compared to other methods. As both \texttt{NBSim} and \texttt{NuQClq} trade time for higher quality solution extraction, it is expected that they may produce slightly higher solutions, i.e., quasi-cliques of larger size. For the two \texttt{NSF} methods, we can observe that their produced solutions are also competitive.
% Notably, our \texttt{DMI-bf} also has achieved the largest average size in some of the data, e.g., rand\_GG. These results prove that our algorithm is competitive in terms of the solution quality. 

Furthermore, we can further see that in rand\_ER, both \texttt{FastNBSim} and \texttt{NSF-fn} achieve a much lower edge-density of $0.74$ compared to others, while \texttt{NBSim} achieves a smaller average size of $9.5$ in rand\_ER. 
On the other hand, none of the results produced by either \texttt{DMI-bt} or \texttt{DMI-bf} are significantly worse than others. This prove the robustness of our proposed algorithm. 

% For comparing \texttt{DMI-bf} and \texttt{DMI-bt}, \texttt{DMI-bf} usually identifies slightly larger subgraphs than \texttt{DMI-bt} while the average edge-density results are very similar. The reason within could be that the bottom-$k$ method produces a slightly more inaccurate estimation than the $l$-buffered method, which is also a trade-off between efficiency and quality: The $l$-buffered method usually takes $O(k \log |A|)$ to update, while the bottom-$k$ only takes $O(k+\log |A|)$ for updates. But in general, both \texttt{DMI-bf} and \texttt{DMI-bt} are strong in solution quality.

% \jb{\textbf{} on the best result, and correct all algorithms name, along with texttt}
% We find that our method can achieve quasi-cliques with comparable or even better sizes and edge densities than the baselines. \jb{more details}

% \todo{2 tables and 2 graphs}
\vspace{0.005cm}
\subsection{Comparison on Special Dynamic Graphs} 
The second set of experiments evaluates the algorithms on datasets in special dynamic settings, including incremental, decremental, and temporal graphs with or without end times.

\begin{table}[t]
\centering
\caption{Comparison of the quasi-clique density and size across algorithms on datasets generated from purely insertions/deletions and temporal graphs.}
\label{tab:other_types_results_transposed}
\scriptsize
\setlength{\tabcolsep}{1pt}
\renewcommand{\arraystretch}{1.1}
\resizebox{\columnwidth}{!}{%
\begin{tabular}{l|cc|cc|cc|cc|cc|cc}
\hline
 & \multicolumn{2}{c|}{del\_FB}
 & \multicolumn{2}{c|}{del\_ER}
 & \multicolumn{2}{c|}{inc\_FB}
 & \multicolumn{2}{c|}{inc\_ER}
 & \multicolumn{2}{c|}{temp\_DB}
 & \multicolumn{2}{c}{tinc\_DB} \\
\cline{2-13}
\textbf{Algorithm}
 & Size & Dens. & Size & Dens. & Size & Dens. & Size & Dens. & Size & Dens. & Size & Dens. \\
\hline
\texttt{DMI-bt} & 109.2 & 0.92 & 10.6 & 0.92 & 113.0 & 0.96 & 12.0 & 0.92 & 16.0 & 0.94 & 17.3 & 1.00 \\
\texttt{DMI-bf} & 82.7 & 0.91 & 11.1 & 0.93 & 91.0 & 0.95 & 13.0 & 0.92 & 15.5 & 0.94 & 16.8 & 1.00 \\
\texttt{NSF-fn} & 73.1 & 0.91 & 18.2 & 0.75 & 39.38 & 0.98 & 15.9 & 0.78 & 13.9 & 0.97 & 16.7 & 1.00 \\
\texttt{NSF-ns} & 30.2 & 0.99 & 7.9 & 0.99 & 8.70 & 1.00 & 9.8 & 1.00 & 14.4 & 0.99 & 17.6 & 1.00 \\
\texttt{FastNBSim} & 73.1 & 0.91 & 18.2 & 0.75 & 83.1 & 0.95 & 15.9 & 0.78 & 13.7 & 0.97 & 16.7 & 1.00 \\
\texttt{NBSim} & 30.6 & 0.97 & 9.7 & 0.90 & 29.1 & 0.99 & 9.7 & 0.92 & 14.1 & 0.97 & 17.6 & 0.94 \\
\texttt{NuQClq} & 140.8 & 0.92 & 28.3 & 0.93 & 115.0 & 0.97 & 29.0 & 0.93 & 12.5 & 0.99 & 14.2 & 1.00 \\
\hline
\end{tabular}%
}
\end{table}

\begin{figure}[t]
    \centering
    %\hspace*{-0.53cm} % 向左移动 0.5 cm，可调为 -1cm、-2cm
    \includegraphics[width=0.95\columnwidth,pagebox=cropbox]{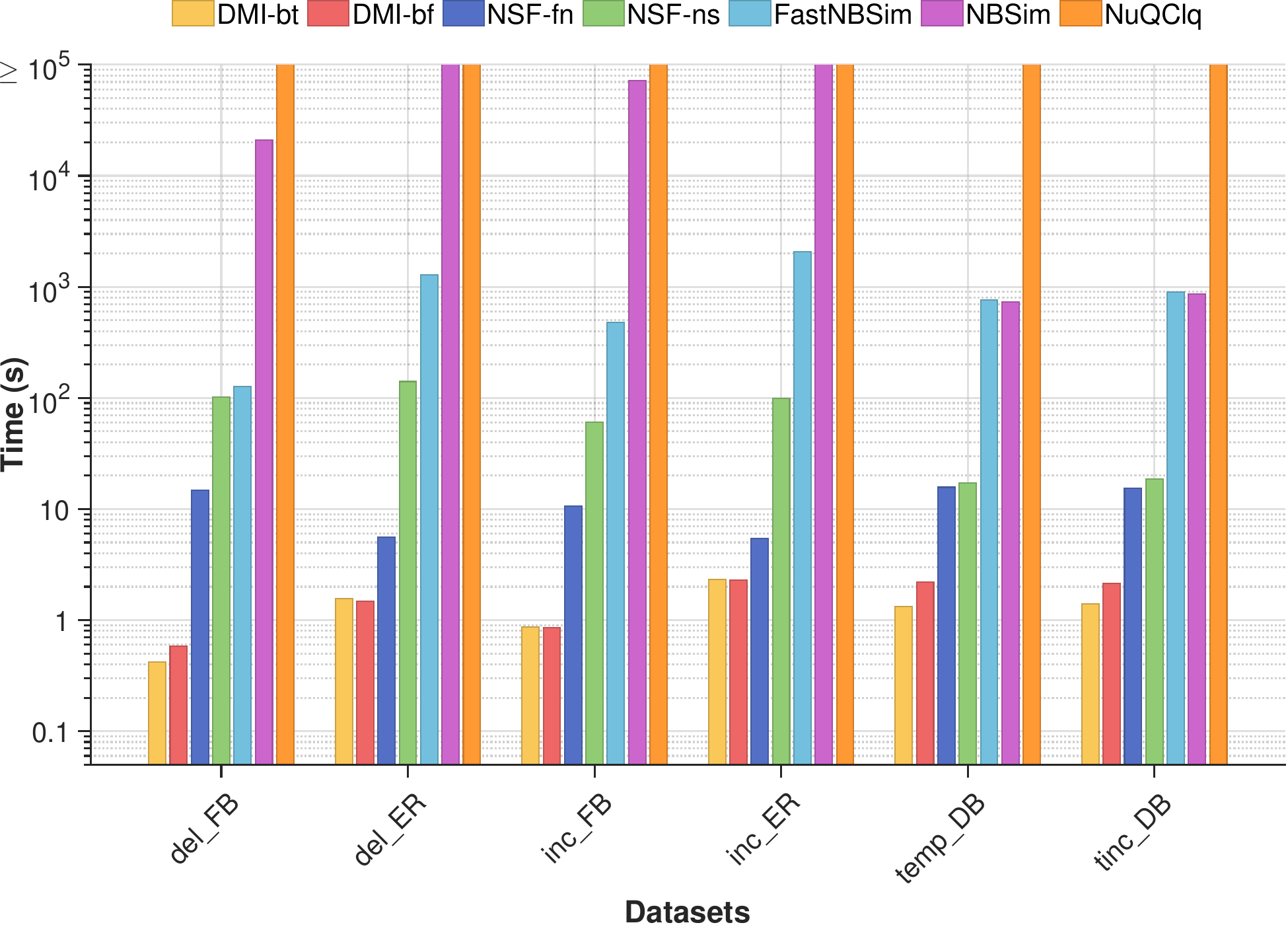} % 放大比例
    \caption{The total runtime (seconds) for all $Q$ updates for different algorithms on datasets generated from purely insertions/deletions and temporal graphs.}
    \label{fig:other_time_result}
\end{figure}

\vspace{0.01cm}
\subsubsection{Efficiency} We compare the total runtime of all algorithms for $Q$ operations, shown in \Cref{fig:other_time_result}. As with the random datasets, both \texttt{DMI-bt} and \texttt{DMI-bf} significantly outperform all other baselines across all datasets and dynamic settings, often by several orders of magnitude.
Across all datasets, \texttt{DMI-bt} and \texttt{DMI-bf} display a total runtime for all $Q$ operations within 10 seconds. In contrast, \texttt{FastNBSim} requires $10^2$ to a few $10^3$ seconds to complete the update sequence. In general, ours give up to a $1,000\times$ improvement over them. For \texttt{NBSim}, it takes over $10^4$ seconds per update sequence for most datasets. \texttt{NuQClq} reaches the $\geq10^5$-second level for every update sequence. 
% Both \texttt{NBSim} and \texttt{NuQClq} are not scalable on these datasets with special dynamic settings.
The efficiency gap holds for incremental and decremental settings, confirming that \texttt{DMI} does not depend on update direction. 
% On the other hand, we can see that the average runtime for the three decremental datasets is lower than the average runtime for the three incremental datasets. Note that they are both generated from FB, CM and ER. This arises because when deletions usually do not touch the current optimal quasi-cliques stored in the candidate list. In this case, our algorithm does not require a massive update. On the other hand, insertions could cause new quasi-cliques to arise, and our algorithm requires a check on each affected neighbor. This could potentially increase the update cost compared to deletion-only datasets.

The results on the temporal graphs confirm the efficiency of \texttt{DMI} on time-dependent updates. Notably, \texttt{NSF-fn} and \texttt{NSF-ns}' performance is approaching \texttt{DMI}, which is expected since these special settings narrow the effect of the edge updates. 
% Runtime stability are also proved by comparing with other baselines' performance. While baselines display strong fluctuations on the different dynamic datasets generated from the same graph, both variants of \texttt{DMI}'s performance remain stable.
% Overall, although the advantage over the state-of-the-art is slightly degraded from $10,000\times$ to $1,000\times$, it is still significant enough.  On the other hand, 
Experiments on special dynamic settings confirm that \texttt{DMI} not only excels on random dynamic graphs but also maintains robust efficiency under insertion-only, deletion-only, and temporal graph updates, further proving its scalability and efficiency in practice.
% In \Cref{fig:rand_xx time outcome},we give the runtime of all algorithms on the same ten datasets. Our proposed method demonstrates a remarkable improvement in computational efficiency. It achieves up to two orders of magnitude speedup over \texttt{FastNBSim} and \texttt{NBSim}, primarily due to its incremental update mechanism that avoids redundant recomputation after each edge modification. Compared with \texttt{NuQClq}, our method is even faster—while maintaining comparable or superior quasi-clique quality.
% It is worth noting that \texttt{NuQClq} did not terminate naturally within the preset time limit in all tested datasets; instead, its runtime reached the maximum allowed threshold, highlighting the substantial computational advantage of our approach. \jb{maybe more number details} This significant efficiency gain stems from our algorithm’s design, which leverages dynamic neighborhood maintenance and effective pruning strategies to minimize unnecessary similarity calculations.
% In contrast, baseline algorithms such as \texttt{NBSim} and \texttt{NuQClq} need to recompute similarities or re-execute full searches for each update, resulting in much higher computational costs.

% \begin{figure}[t]
%     \centering
%     \includegraphics[width=0.5\textwidth]{fig/timcmp.png}
%     \caption{Running Time of each Algorithms}
%     \label{fig:performance}
% \end{figure}

% In Figure 1, we present the efficiency comparison among all tested algorithms.
% \jb{I stop at here, please correct the algorithm name, etc}
\vspace{0.01cm}
\subsubsection{Solution Quality} We compute the average density and size of the detected quasi-clique for all operations, shown in \Cref{tab:other_types_results_transposed}. We can see that \texttt{DMI-bt} and \texttt{DMI-bf} again achieve consistently high densities ($\delta(S)\geq 0.91$) in all datasets. Similarly, the solution size for both algorithms remains competitive with other baselines' results. 
% We can also see that in some of the datasets, e.g., temp\_DB, \texttt{DMI-bt} and \texttt{DMI-bf} give a better average solution size than any other baselines, including \texttt{NuQClq}.
On the other hand, we can observe that \texttt{FastNBSim} displays a notable degradation in solution quality relative to \texttt{DMI}. For example, on del\_ER and inc\_ER, the average edge-density drops to $0.75$ and $0.78$ compared to other results of over $0.9$, while the solution size does not increase much. 

For \texttt{NuQClq}, it still produces the most competitive solutions in the majority of datasets. However, we can see that in some datasets, e.g., tinc\_DB, the produced average solution size is smaller than that of any other algorithm. We can also see that the performance of \texttt{NSF} drops over these special setting, e.g. in del\_FB, inc\_FB. This aligns with our analysis that \texttt{NSF} would fail to account the candidate with great potential over edge updates.

% Compared with the results in the random datasets, the overall solution quality of \texttt{DMI-bt} seems to be better. We can observe that in a number of datasets, the average solution size is larger, such as del\_FB, inc\_FB, and temp\_DB. It also sometimes generates a higher edge-density in average. Despite of these, we can still conclude that the performance between \texttt{DMI-bt} and \texttt{DMI-bf} remains to be similar, which proves the stability of our algorithm framework. In all, from all these graph datasets with special dynamic settings, the advantage of our \texttt{DMI} in terms of the solution size is displayed even more obviously.

\section{ADDITIONAL EXPERIMENT}
\label{sec:furtherexperiment}
\subsection{Effect of Parameters}
In this section, we study the effect of different hyperparameters, including the following:
\begin{enumerate}[leftmargin=*]
    \item $\gamma,b$: These two parameters together determine the overall solution quality of the extracted quasi-clique as well as the runtime.
    \item $batch$: In both our insert and delete routine, this parameter decides the massive rebuild frequency, indicating calling a \texttt{Build} for every $batch$ operations.
    \item $k$: Here, $k$ refers to both the number of independent hash functions in the $l$-buffered-$k$ method and the number of smallest elements in the Bottom-$k$ method.
\end{enumerate}

We conduct an ablation study on $B$, i.e., the number of quasi-cliques kept in the candidate list $L$ in \Cref{sec:abalation}.

\begin{figure}[t]
    \centering
    \hspace*{-0.23cm} % 向左移动 0.5 cm，可调为 -1cm、-2cm
    \includegraphics[width=0.92\columnwidth,pagebox=cropbox]{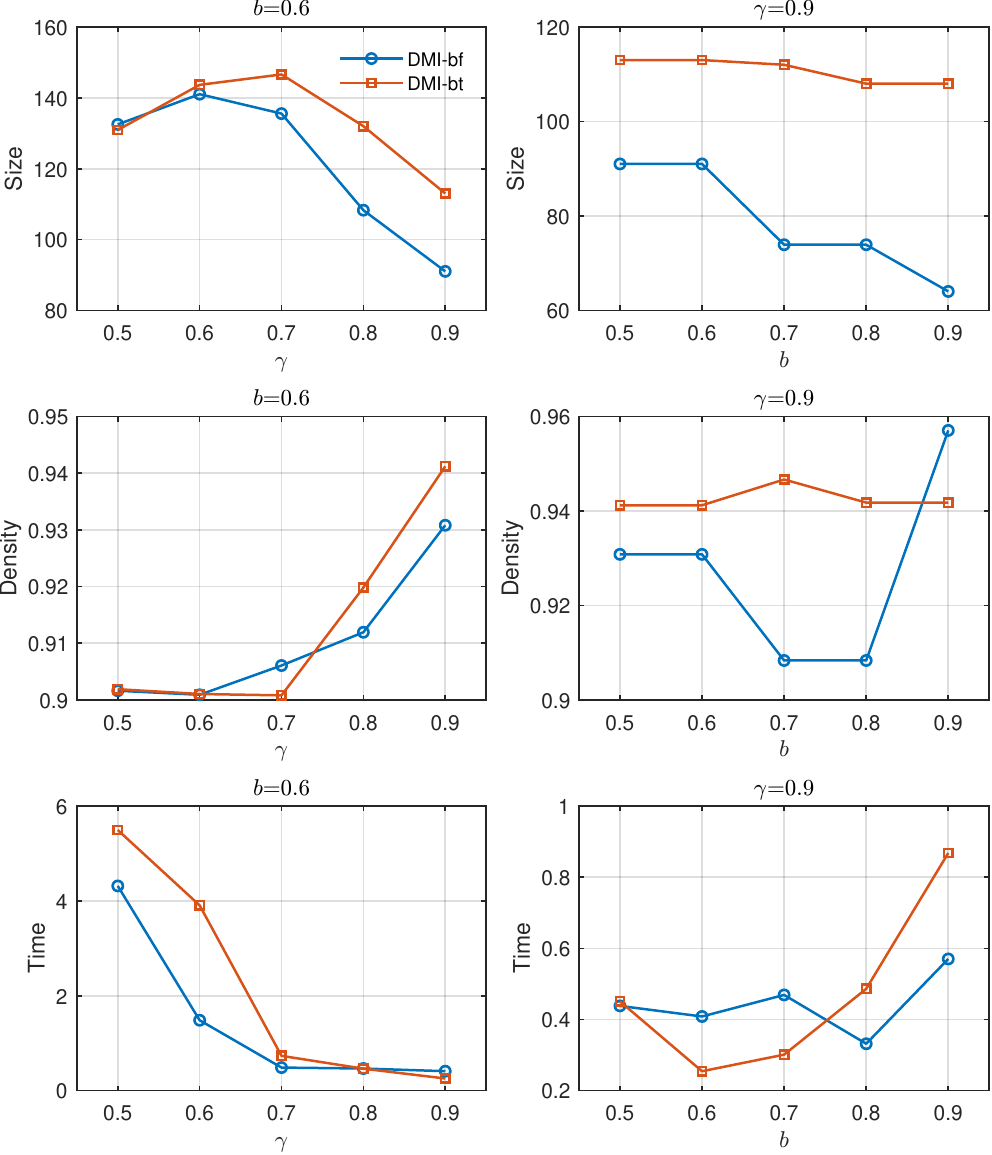}
    \caption{The total runtime (seconds), average edge-density, and average size of \texttt{DMI-bf} and \texttt{DMI-bt} for different $\gamma$ and $b$ on rand\_FB.}
    \label{fig:gamma and b}
\end{figure}

\paragraph{Effect of $\gamma$ and $b$} We fix $b=0.6$ for $\gamma \in [0.5,0.9]$ as well as fixing $\gamma=0.9$ for $b \in [0.5,0.9]$. All experiments are conducted on rand\_FB and are shown in \Cref{fig:gamma and b}. 

We can observe that in terms of the quasi-clique size, increasing $b$ causes a decrease, while increasing $\gamma$ first raises the size and then reduces it after $\gamma=0.6$ for \texttt{DMI-bf} and $0.7$ for \texttt{DMI-bt}. $\gamma$ determines the threshold for selecting any neighbor. A larger $\gamma$ makes it harder to form a large quasi-clique. This matches our observation. On the other hand, since $b$ decides whether the current subgraph extracted from the neighborhood is admitted, a larger $b$ also makes it harder to generate a large one. From the experimental results, we can see that $b$ is relatively moderate than $\gamma$.

While fixing $b$ and increasing $\gamma$, we can see the edge-density is increasing for both methods. This is as expected since $\gamma$ directly decides whether any vertex can be admitted into the quasi-clique. On the other hand, for the case of increasing $b$ while fixing $\gamma$, its effect is less obvious: For \texttt{DMI-bt}, there is almost no difference. We conjecture that it is mainly because it is only used to reject quasi-cliques after extracting from the neighborhood, making it less direct for the solution.

Now we discuss the total runtime. We can see that as $\gamma$ increases, the runtime decreases dramatically. Since $\gamma$ is used to reject vertices with lower potentials, a larger $\gamma$ will cause fewer vertices to be included, decreasing the overall time cost when updates are made. On the other hand, increasing $b$ seems to significantly increase the runtime overall, with fluctuations. This may be because a larger $b$ makes it easier to reject quasi-cliques just been extracted, thus we have to search for more neighborhoods to fill up the candidate list, increasing runtime.

% From Figure~2, the quasi-clique size, density, and running time under different $\gamma$ and $b$ values are compared for DMI-bf and DMI-bt.
% When $\gamma$ increases with a fixed $b=0.6$ (left column), the quasi-clique size gradually decreases, while its density slightly rises, indicating that a larger $\gamma$ enforces stricter inclusion thresholds and yields denser but smaller clusters. 
% Meanwhile, the extraction time also increases due to the higher computational cost of filtering and validation under stronger constraints. 
% Conversely, when $b$ increases with $\gamma$ fixed at $0.9$ (right column), a similar trend is observed: the size shrinks and density improves for $b$ between 0.6 and 0.9.
% Moreover,

\begin{figure}[t]
    \centering
    \hspace*{-0.23cm} % 向左移动 0.5 cm，可调为 -1cm、-2cm
    \includegraphics[width=0.92\columnwidth,pagebox=cropbox]{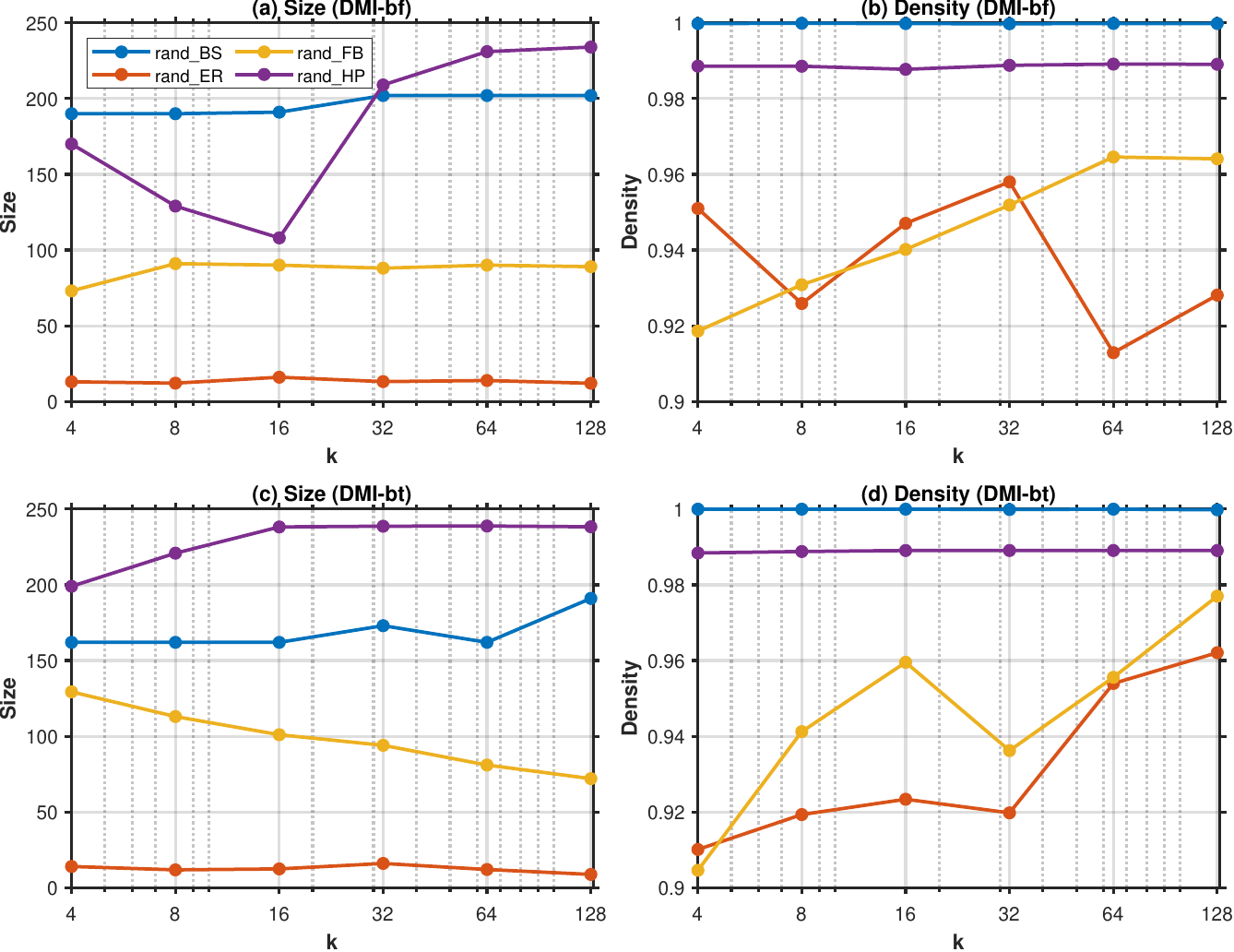}
    \caption{The average size and density of \texttt{DMI-bf} and \texttt{DMI-bt} for different $k$ on rand\_BS, rand\_FB, rand\_ER, rand\_HP.}
    \label{fig:k}
\end{figure}

\paragraph{Effect of varying $k$} We conduct experiments to compare the performance of \texttt{DMI-bf} and \texttt{DMI-bt} on rand\_FB, rand\_HP, rand\_BS, and rand\_ER, where the parameter $k$ varies from $4$ to $128$, with $\gamma$ fixed at $0.9$ and $b$ fixed at $0.6$. Overall, both methods produce close and high-quality results. The results are illustrated in \Cref{fig:k}.

For \texttt{DMI-bf} that utilizes the $l$-buffered-$k$-MinHash method, we can see that the performance is either converging or fluctuating for the different values of $k$. That is to say, increasing the number of distinct hash values generated does not have a significant impact on the solution quality for most datasets, except for the size on rand\_HP, both in terms of size and density. This matches with the observation from \cite{pang2024similarity}, which uses a similar hashing method, saying that a small number of $k$ is sufficient enough to generate high-quality solutions.

On the other hand, the impact of increasing $k$ is relatively more obvious to be observed for \texttt{DMI-bt}, which uses the Bottom-$k$-MinHash method. We can see that on rand\_HP, the average solution size increases when $k$ gets larger, and the density gets larger as well for both rand\_FB and rand\_ER. Although these patterns are still less promising, we can still conjecture that the different mechanism of the hashing method is the main reason for causing such a difference. As in theory, as $k$ gets larger, the estimation of the Jaccard similarity should be more accurate, which should in turn result in an increase of the solution quality. 

% When k is small, approximation errors become more evident — the detected quasi-cliques tend to be larger with lower edge densities, especially in the FB dataset. These inaccuracies stem from the MinHash-based similarity estimation, which may overestimate or underestimate vertex similarities for small k, as also observed in the GG and BS datasets when k = 4. As k increases, the approximation becomes more accurate, and the results gradually converge to those of the baseline.

\begin{figure}[t]
    \centering
    \hspace*{-0.28cm} % 向左移动 0.5 cm，可调为 -1cm、-2cm
    \includegraphics[width=0.92\columnwidth,pagebox=cropbox]{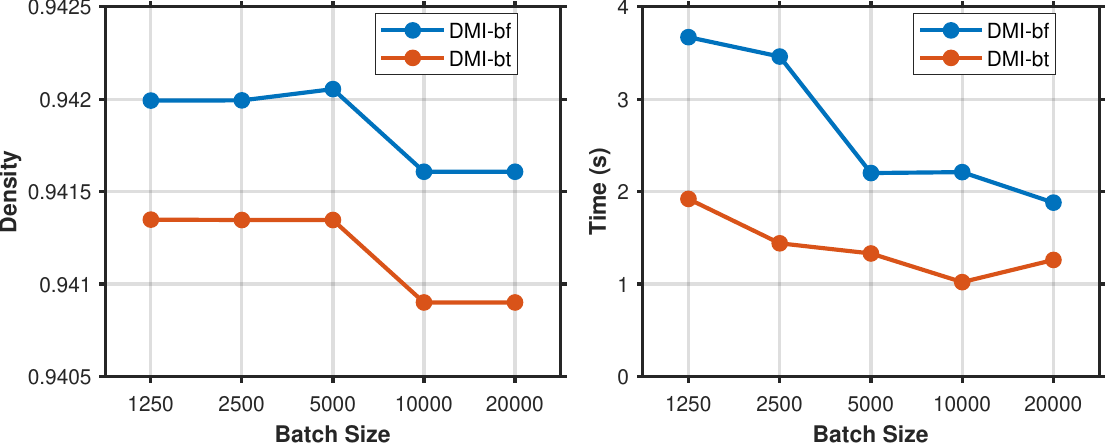}
    \caption{The average density and total runtime (seconds) of \texttt{DMI-bf} and \texttt{DMI-bt} for different $batch$ on Graph temp\_DB.}
    \label{fig:batch}
\end{figure}

\paragraph{Effect of varying $batch$} We conduct experiments to compare the effect on different $batch$, the parameter that decides the rebuild frequency. Specifically, we examined the total runtime and the edge-density for both \texttt{DMI-bf} and \texttt{DMI-bt} on temp\_DB. The results are shown in \Cref{fig:batch}.

We can observe that as $batch$ increases, the solution quality drops a little while still keeping a good result. Since it becomes less frequent to rebuild the whole candidate list, falling into bias could reduce solution quality. But in general, as we keep a list of candidates instead of one, such bias will not be significant as shown in \Cref{fig:batch} for both methods.

In terms of the total runtime, as expected, since each rebuild takes a large runtime, less rounds of rebuild should decrease the runtime in general, which is observed in \Cref{fig:batch} as well.

% As illustrated in Figure~4, the elapsed time of both DMI-bf and DMI-bt decreases steadily with increasing batch size. The trend is approximately linear with minor fluctuations. 

% In contrast, the resulting edge density exhibits a slightly decreasing tendency as the batch size grows, although this trend is not strictly monotonic. The small deviations observed suggest that larger batch sizes may introduce marginal differences in approximation or update ordering, leading to subtle variations in the final quasi-clique density.

\subsection{Scalability on Operation Size}
\begin{figure}[t]
    \centering
    \hspace*{-0.4cm} % 向左移动 0.5 cm，可调为 -1cm、-2cm
    \includegraphics[width=0.95\columnwidth,pagebox=cropbox]{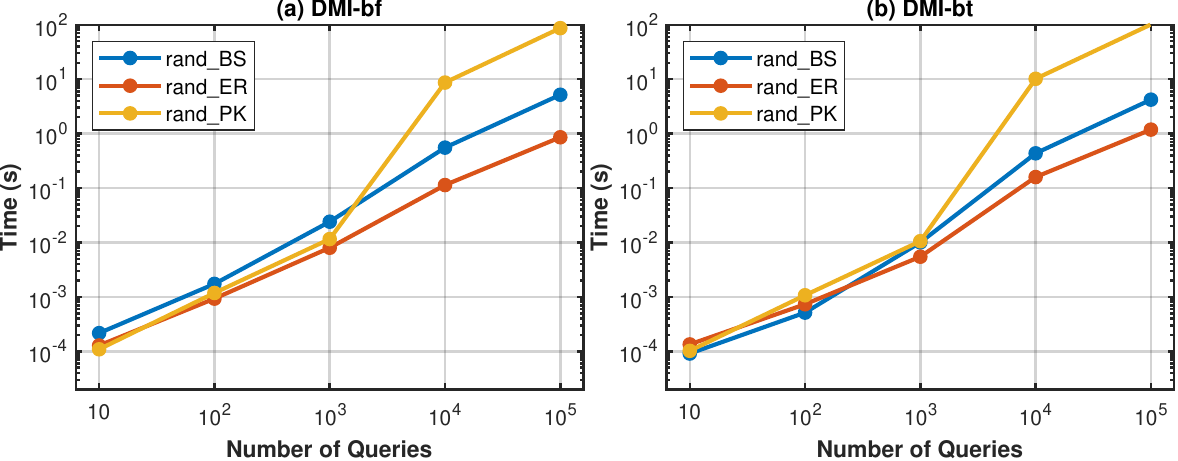}
    % \hspace*{-0.4cm} % 向左移动 0.5 cm，可调为 -1cm、-2cm
    % \vspace{0.2cm}
    % \includegraphics[width=0.48\textwidth]{fig/图3.png}
    \caption{The Running Time of \texttt{DMI-bf} and \texttt{DMI-bt} for increasing Q on Graph rand\_BS, rand\_ER and rand\_PK.}
    \label{fig:Q}
\end{figure}

\begin{figure}[t]
    \centering
    \hspace*{-0.4cm} % 向左移动 0.5 cm，可调为 -1cm、-2cm
    \includegraphics[width=0.95\columnwidth,pagebox=cropbox]{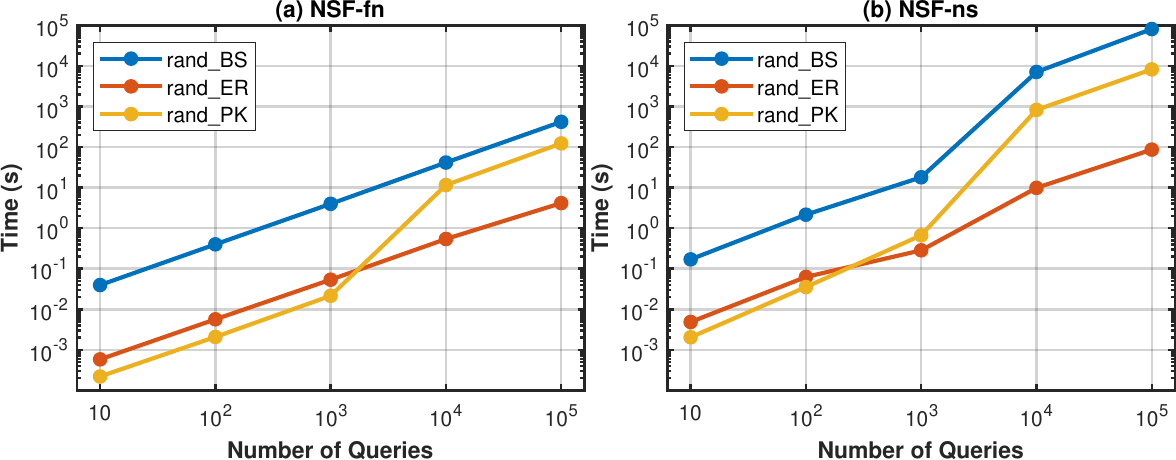}
    \caption{The Running Time of \texttt{NSF-fn} and \texttt{NSF-ns} for increasing Q on Graph rand\_BS, rand\_ER and rand\_PK.}
    \label{fig:Q_NSF}
\end{figure}

We conduct an independent experiment to further demonstrate the scalability of our algorithm. On three datasets rand\_BS, rand\_ER, rand\_PK, we generate $10,10^2,10^3,10^4,10^5$ operations and compare the \textbf{total} runtime for \texttt{DMI-bf}, \texttt{DMI-bt}, \texttt{NSF-fn} and \texttt{NSF-ns}. The results are shown in \Cref{fig:Q} and \Cref{fig:Q_NSF}.

We can observe that the total runtime of all four algorithms increases with $Q$. This observation aligns well with our analysis, as the computational cost grows proportionally to the number of affected vertices and edges. With a small number of operations, the original graph structure is only mildly changed; thus, the optimal quasi-clique is almost unaffected. However, as the number of operations increases, the structure may be significantly impacted, triggering more extraction and rebuilding related to the updated edge and its neighborhood. On the other hand, once the whole structure is altered, it is likely to stay stable for some time. Therefore, for real-world datasets, it is expected to follow a linear trend as long as our algorithm ensures the time cost is proportional to the amount of changes.
Moreover, we can also see that the curves of \texttt{DMI-bf} and \texttt{DMI-bt} are similar to each other, and they both have a slightly steeper increase from $10^3$ to $10^4$. The closeness also indicates the stability of our algorithm framework.

% \todo{Analysis of Running Time of \texttt{NSF-fn} and \texttt{NSF-ns} for increasing Q}

\subsection{Time Cost of Different Stages}

% %原本没有NSF的表
% \begin{table}[t]
% \centering
% \small
% \caption{Initialization Time and Update Time Comparison}
% \vspace{0.2cm}
% \label{tab:init and query}
% \begin{tabular}{lcccc}
% \hline
%  & \multicolumn{2}{c}{\texttt{DMI-bf}} & \multicolumn{2}{c}{\texttt{DMI-bt}} \\
% \cline{2-5}
% Dataset & Init Time & Update Time & Init Time & Update Time \\
% \hline
% del\_BS & 0.36 & 4.74 & 0.39 & 3.71 \\
% del\_CM & 0.02 & 0.43 & 0.01 & 0.14 \\
% rand\_ER & 0.23 & 1.02 & 0.11 & 1.38 \\
% rand\_GG & 0.20 & 4.39 & 0.24 & 3.39 \\
% inc\_BS & 0.39 & 5.59 & 0.40 & 4.31 \\
% inc\_FB & 0.04 & 0.45 & 0.02 & 0.46 \\
% % temp\_FW & 0.69 & 2.22 & 0.43 & 4.45 \\
% % temp\_YG & 2.70 & 12.12 & 1.58 & 16.5 \\
% % tinc\_FW & 0.68 & 2.46 & 0.38 & 4.83 \\
% % tinc\_YG & 2.47 & 10.17 & 1.23 & 12.47 \\
% \hline
% \end{tabular}
% \vspace{0.2cm}
% \end{table}

% DMI values use the corrected binary. NSF values are unchanged.
\begin{table}[t]
\centering
\setlength{\tabcolsep}{1pt}
\caption{Initialization Time and Total Update Time Comparison (seconds)}
\label{tab:init and query}
\scriptsize
\resizebox{\columnwidth}{!}{%
\begin{tabular}{lcccccccc}
\hline
 & \multicolumn{2}{c}{\texttt{DMI-bf}} & \multicolumn{2}{c}{\texttt{DMI-bt}}
 & \multicolumn{2}{c}{\texttt{NSF-fn}} & \multicolumn{2}{c}{\texttt{NSF-ns}} \\
\cline{2-9}
Dataset & Init & Update & Init & Update & Init & Update & Init & Update \\
\hline
del\_BS  & 0.36 &  4.83 &  0.41 &  3.87 & 1.17 & 528.10 & 3411 & 68060 \\
del\_CM  & 0.02 &  0.60 & 0.01 &  0.22 & 0.03 & 1.02 & 0.41 & 8.54 \\
rand\_ER &  0.21 &  0.86 & 0.11 &  1.19 & 0.04 & 5.07 & 3.64 & 107.6 \\
rand\_GG &  0.19 &  4.09 &  0.22 &  3.15 & 1.39 & 35.76 & 125.8 & 2547 \\
inc\_BS  &  0.37 &  6.67 & 0.40 &  4.42 & 1.33 & 537.70 & 3429 & 87810 \\
inc\_FB  & 0.04 &  0.68 & 0.02 & 0.68 & 0.01 & 10.55 & 1.35 & 59.28 \\
\hline
\end{tabular}%
}
\end{table}

% \begin{table}[t]
% \centering
% \small
% \caption{Initialization Time and Update Time Comparison}
% \vspace{0.2cm}
% \setlength{\tabcolsep}{2pt}
% \label{tab:init and query}
% \begin{tabular}{lcccccccc}
% \hline
%  & \multicolumn{2}{c}{\texttt{DMI-bf}} & \multicolumn{2}{c}{\texttt{DMI-bt}}& \multicolumn{2}{c}{\texttt{NSF-fn}} & \multicolumn{2}{c}{\texttt{NSF-ns}} \\
% \cline{2-9}
% Dataset & Init & Update & Init & Update & Init & Update & Init & Update \\
% \hline
% del\_BS & 0.36 & 4.74 & 0.39 & 3.71 & 1.17 & 528.10 & 3411 & 68060\\
% del\_CM & 0.02 & 0.43 & 0.01 & 0.14 & 0.03 & 1.02 & 0.41 & 8.54\\
% rand\_ER & 0.23 & 1.02 & 0.11 & 1.38 & 0.04 & 5.07 & 3.64 & 107.6\\
% rand\_GG & 0.20 & 4.39 & 0.24 & 3.39 & 1.39 & 35.76 & 125.8 & 2547\\
% inc\_BS & 0.39 & 5.59 & 0.40 & 4.31 & 1.33 & 537.70 & 3429 & 87810\\
% inc\_FB & 0.04 & 0.45 & 0.02 & 0.46 & 0.01 & 10.55 & 1.35 & 59.28\\
% temp\_FW & 0.69 & 2.22 & 0.43 & 4.45 & 0.12 & 4.44 & 6.45 & 191\\
% temp\_YG & 2.70 & 12.12 & 1.58 & 16.5 & 2.87 & 84.96 & 197.5 & 53650\\
% tinc\_FW & 0.68 & 2.46 & 0.38 & 4.83 & 0.12 & 4.95 & 6.35 & 194.6\\
% tinc\_YG & 2.47 & 10.17 & 1.23 & 12.47 & 3.22 & 91.39 & 197.4 & 17410\\
% \hline
% \end{tabular}
% \vspace{0.2cm}
% \end{table}

We conduct an experiment to evaluate the runtime for both the initialization (\texttt{Rebuild}) and edge updates (\texttt{AddEdge} and \texttt{DeleteEdge}). Specifically, we select six different datasets of the three different dynamic settings. The runtimes (in seconds) are reported in \Cref{tab:init and query}.

As we can see, for both \texttt{DMI-bf} and \texttt{DMI-bt}, the initialization time is significantly smaller than the total update time cost. On the other hand, \texttt{DMI-bt} has faster initialization on three of the six datasets and lower total update times on four.

From the result of \texttt{NSF-fn} and \texttt{NSF-ns}, we can see that their initialization time is usually larger then both \texttt{DMI-bf} and \texttt{bt}, as they require a neighbor search for every vertex, without any pruning. We can also observe that the update time is also large and varies a lot between different datasets. \texttt{NSF} does not exploit quasi-clique structure, whereas \texttt{DMI} does.

\subsection{Ablation Study}
\label{sec:abalation}

% Verified with the corrected binary; displayed values are unchanged.
\begin{table}[t]
\centering
\small
\caption{The improvement of average size from B=1 to B=9}
\label{tab:B=1 or 9}
\begin{tabular}{lcc}
\hline
Dataset & $B=1$ & $B=9$ \\
\hline
inc\_HP  & 92.00 & 129.00 \\
rand\_ER & 11.99 & 12.09 \\
rand\_HP & 92.00 & 129.00 \\
del\_CM  & 21.62 & 22.39 \\
del\_ER  & 10.93 & 11.09 \\
del\_FB  & 82.58 & 82.89 \\
del\_HP  & 83.81 & 128.57 \\
\hline
\end{tabular}
\end{table}

\begin{figure}[t]
    \centering
    \hspace*{-0.33cm} % 向左移动 0.5 cm，可调为 -1cm、-2cm
    \includegraphics[width=0.86\columnwidth,pagebox=cropbox]{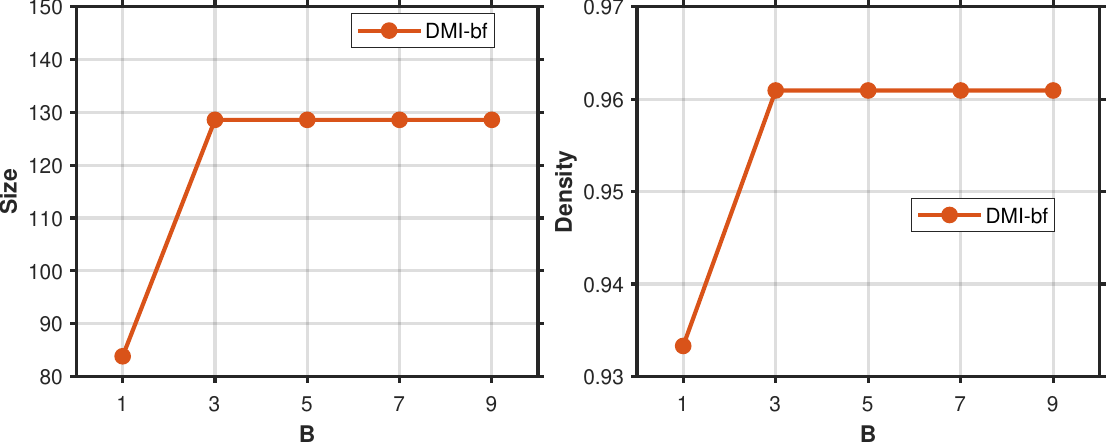}
    \caption{The average size and density of \texttt{DMI-bf} for different $B$ on Graph del\_HP.}
    \label{fig:B}
\end{figure}

One of the key parts of our algorithm framework is maintaining a number of candidate quasi-cliques in the list instead of one. To evaluate the effect of maintaining such a candidate list, we conduct an ablation study by mainly comparing the results under 
$B=1$ and $B=9$. In addition, we also show the result of varying $B$ for the dataset del\_HP, in terms of both the size and the edge-density (\Cref{fig:B}).

To begin with, we report the detailed improvement of the average quasi-clique size between $B=1$ and $B=9$ for \texttt{DMI-bf} in \Cref{tab:B=1 or 9}. We select some datasets from each of the dynamic settings, and we can see that there is a significant improvement over the average size for $B=1$. This aligns with our expectation: as $B$ increases, the algorithm is allowed to explore a broader search space, resulting in the growth of the number of candidate solutions, enabling the algorithm to locate quasi-cliques with either high quality or good potential. On the contrary, if we only keep the current "optimal" quasi-clique, as the graph evolves dynamically, it is easy to become biased and thus miss finding better solutions. This proves the necessity and validates the effectiveness of the candidate list.

\subsection{Memory Usage}

% \begin{table}[t]
% \centering
% \caption{Memory(kbytes)}
% \label{tab:rand_memoryuse_results}
% \small
% \setlength{\tabcolsep}{0.8pt}
% \renewcommand{\arraystretch}{1.2}
% \begin{tabular}{l|cc|cc|cc}
% \hline
%   &\multicolumn{2}{c|}{\textbf{k=8}} &\multicolumn{2}{c|}{\textbf{k=16}} & \multicolumn{2}{c}{\textbf{k=32}} \\ 
% \cline{2-7}
% \textbf{Dataset} & DMI-bf & DMI-bt & DMI-bf & DMI-bt & DMI-bf & DMI-bt \\ 
% \hline
% rand\_BS & 1724004 & 1181368 & 2446668 & 1224500 & 3889368 & 1309992  \\
% rand\_CM   & 81708 & 34540 & 140608 & 35940 & 258644 & 39096 \\
% rand\_ER   & 121432 & 62356 & 203896 & 64476 & 368812 & 69224 \\
% rand\_FB   & 30856 & 27988 & 42792 & 27992 & 68328 & 28712 \\
% rand\_GG   & 1527024 & 857408 & 2366508 & 912028 & 4046620 & 1021484 \\
% rand\_GW   & 535924 & 265116 & 911188 & 275800 & 1668560 & 300932 \\
% rand\_HP   & 55596 & 37516 & 86484 & 38376 & 148836 & 39880  \\
% rand\_PK   & 7216732 & 5219720 & 11531872 & 5321592 & 19170704 & 5524172 \\
% rand\_SF   & 762424 & 417632 & 1214024 & 428112 & 2130552 & 463436 \\
% rand\_TC   & 8180396 & 6227012 & 12810736 & 6380176 & 22084384 & 6575624 \\
% \hline
% \end{tabular}
% \end{table}

% Peak RSS from /usr/bin/time -v, converted from KiB to MiB by dividing by 1024.
\begin{table}[t]
\centering
\caption{Peak Memory Usage of \texttt{DMI-bf} and \texttt{DMI-bt} (MiB)}
\label{tab:rand_memoryuse_results}
\small
\setlength{\tabcolsep}{0.8pt}
\renewcommand{\arraystretch}{1.05}
\begin{tabular}{l|cc|cc|cc}
\hline
 & \multicolumn{2}{c|}{$k=8$} & \multicolumn{2}{c|}{$k=16$} & \multicolumn{2}{c}{$k=32$} \\
\cline{2-7}
\textbf{Dataset} & \texttt{DMI-bf} & \texttt{DMI-bt} & \texttt{DMI-bf} & \texttt{DMI-bt} & \texttt{DMI-bf} & \texttt{DMI-bt} \\
\hline
rand\_BS & 1684.88 & 1155.38 & 2390.87 & 1197.21 & 3798.95 & 1280.84 \\
rand\_CM & 80.25 & 34.34 & 137.54 & 35.48 & 252.75 & 38.72 \\
rand\_ER & 119.36 & 61.73 & 199.80 & 63.88 & 361.16 & 68.36 \\
rand\_FB & 30.59 & 27.94 & 42.33 & 28.04 & 67.45 & 28.39 \\
rand\_GG & 1492.26 & 838.23 & 2311.87 & 891.86 & 3952.49 & 998.23 \\
rand\_HP & 54.80 & 37.38 & 85.10 & 37.88 & 146.56 & 39.38 \\
rand\_PK & 7048.77 & 5096.77 & 11263.20 & 5196.59 & 18724.15 & 5394.89 \\
rand\_SF & 745.21 & 408.81 & 1186.22 & 419.11 & 2081.61 & 453.41 \\
\hline
\end{tabular}
\end{table}

% Source files report maximum RSS in KiB. Values below are KiB / 1024 = MiB.
\begin{table}[t]
\centering
\caption{Peak Memory Usage of \texttt{NSF-fn} and \texttt{NSF-ns} for $k=8$ (MiB)}
\label{tab:NSF_rand_memoryuse_results}
\small
\setlength{\tabcolsep}{1pt}
\renewcommand{\arraystretch}{1.05}
\begin{tabular}{l|cc}
\hline
\textbf{Dataset} & \texttt{NSF-fn} & \texttt{NSF-ns} \\
\hline
rand\_BS & 195.08 & 151.26 \\
rand\_CM & 8.69 & 7.26 \\
rand\_ER & 12.31 & 9.65 \\
rand\_FB & 5.54 & 5.17 \\
rand\_GG & 205.69 & 145.80 \\
rand\_HP & 6.95 & 6.32 \\
rand\_PK & 539.86 & 427.35 \\
rand\_SF & 74.82 & 57.04 \\
\hline
\end{tabular}
\end{table}

% Rebuild share and per-update latency from corrected DMI-bf runs.
\begin{table}[!t]
\centering
\caption{Update latency (seconds) for \texttt{DMI-bf}.}
\label{tab:latency-bf}
\small
\setlength{\tabcolsep}{5pt}
\renewcommand{\arraystretch}{1.12}
\begin{tabular}{l|ccc}
\hline
\textbf{Dataset} & \textbf{Rebuild (\%)} & \textbf{Median} & \textbf{Maximum} \\
\hline
rand\_FB & 10.54 & $3\times10^{-6}$   & $4\times10^{-3}$ \\
rand\_BS & 69.96 & $1.2\times10^{-5}$ & $2\times10^{-1}$ \\
rand\_HP & 18.46 & $4\times10^{-6}$   & $8\times10^{-3}$ \\
rand\_ER & 39.49 & $4\times10^{-6}$   & $2\times10^{-2}$ \\
rand\_CM & 8.22  & $4\times10^{-6}$   & $4\times10^{-3}$ \\
temp\_DB & 22.56 & $3\times10^{-6}$   & $8\times10^{-2}$ \\
\hline
\end{tabular}
\end{table}

% Rebuild share and per-update latency from corrected DMI-bt runs.
\begin{table}[!t]
\centering
\caption{Update latency (seconds) for \texttt{DMI-bt}.}
\label{tab:latency-bt}
\small
\setlength{\tabcolsep}{5pt}
\renewcommand{\arraystretch}{1.12}
\begin{tabular}{l|ccc}
\hline
\textbf{Dataset} & \textbf{Rebuild (\%)} & \textbf{Median} & \textbf{Maximum} \\
\hline
rand\_FB & 19.70 & $2\times10^{-6}$ & $5\times10^{-3}$ \\
rand\_BS & 88.84 & $3\times10^{-6}$ & $2\times10^{-1}$ \\
rand\_HP & 48.06 & $2\times10^{-6}$ & $3\times10^{-2}$ \\
rand\_ER & 69.89 & $2\times10^{-6}$ & $6\times10^{-2}$ \\
rand\_CM & 20.29 & $2\times10^{-6}$ & $4\times10^{-3}$ \\
temp\_DB & 24.72 & $2\times10^{-6}$ & $3\times10^{-2}$ \\
\hline
\end{tabular}
\end{table}

We report the peak memory usage (MiB) for both \texttt{DMI-bf} and \texttt{DMI-bt} over eight datasets, shown in \Cref{tab:rand_memoryuse_results}. We can see that the total memory generally grows with the number of vertices and edges. As $k$ increases, the memory also increases in a linear manner. Between the two methods, \texttt{DMI-bf} in general takes up more space as it has a buffer structure. On the other hand, from \Cref{tab:NSF_rand_memoryuse_results}, we can see that for the $k=8$ case, both \texttt{NSF-fn} and \texttt{NSF-ns} require a smaller memory to execute than either \texttt{DMI} variant on all eight datasets.

\subsection{Rebuild Time Cost}

% Q1: The method relies on periodic rebuilding after a fixed number of updates. Can the authors report the rebuild overhead more clearly, such as the fraction of total time spent on rebuilding and the worst-case or percentile update latency? The current paper mainly reports average behavior.

\Cref{tab:latency-bf,tab:latency-bt} report additional rebuild-overhead and update-latency statistics. The rebuild share ranges from 8.22\% to 69.96\% for \texttt{DMI-bf} and from 19.70\% to 88.84\% for \texttt{DMI-bt}; these higher percentages mainly reflect the extremely low cost of ordinary updates (\(10^{-6}\)-\(10^{-5}\) seconds), which makes rebuild time appear larger in proportion. Update latency remains consistently low, with median latencies ranging from $2\times10^{-6}$ to $1.2\times10^{-5}$ seconds and the worst cases dominated by rebuilds.

\subsection{Discard and Rebuild Statistics}

% Q4: Under deletions, the method only tries one-vertex or two-vertex repair. Otherwise, it discards the quasi-clique as a significant change. Can the authors report how often this discard case happens, and how much it affects final solution quality?

\Cref{tab:discard} reports discard and rebuild statistics in the format of \texttt{DMI-bf}/\texttt{DMI-bt} with $B=5$; the discard case happens extremely rarely across all datasets, indicating that the 1- or 2-vertex repair strategy is sufficient in almost all tested repair cases.

% Discard is shown as DMI-bf/DMI-bt. Rebuild counts are equal for both variants.
\begin{table}[!t]
\centering
\caption{Discard and rebuild statistics.}
\label{tab:discard}
\small
\setlength{\tabcolsep}{2pt}
\renewcommand{\arraystretch}{1.20}
\begin{tabular}{l|r|c|r}
\hline
\textbf{Dataset} & \textbf{Updates} & \textbf{Discard (bf/bt)} & \textbf{Rebuild} \\
\hline
del\_FB & 58822 & 4/0 & 11 \\
del\_CM & 62292 & 3/0 & 12 \\
del\_ER & 100000 & 3/5 & 20 \\
del\_SF, del\_BS, del\_GG & 100000 & 0/0 & 20 \\
temp\_DB & 100000 & 4/1 & 20 \\
\hline
Total & 621114 & 14/6 & 123 \\
\hline
\end{tabular}
\end{table}

\section{CONCLUSION}
\label{sec:conclude}

In this paper, we studied the Dynamic Maximum Quasi-Clique Problem (DMQCP), which aims to efficiently maintain the largest $\alpha$-quasi-clique under continuous graph updates. To address the challenges of scalability and update efficiency, we proposed \texttt{DMI}, a novel MinHash-based framework that incrementally maintains candidate quasi-cliques with high accuracy. By integrating $l$-buffered $k$-MinHash and Bottom-$k$ MinHash with a batch reconstruction strategy, \texttt{DMI} effectively balances solution quality and computational cost. As a side product, we also propose a framework \texttt{NSF} that primarily uses the neighbor-search routine to accommodate edge updates.

Extensive experiments on both real-world and synthetic datasets show that \texttt{DMI} achieves up to four orders of magnitude speedup over static baselines while preserving solution quality, enabling real-time dense subgraph mining in evolving networks. Our study opens new directions for dynamic dense subgraph analysis, with potential extensions to other relaxed dense structures (e.g., $k$-plexes, $\delta$-cliques) and more general streaming or distributed settings.

\section*{Acknowledgment}
Chenhao Ma was partially supported by NSFC under Grant 62672430, Basic and Applied Basic Research Fund in Guangdong Province under Grant 2025A1515010439,  Ant Group through CCF-Ant Research Fund, 1+1+1 CUHK-CUHK(SZ)-GDSTC Joint Collaboration Fund - Grant 2025A0505000079, and the Guangdong Provincial Key Laboratory of Big Data Computing, The Chinese University of Hong Kong, Shenzhen.

\section*{AI-Generated Content Acknowledgment}
The authors acknowledge the use of ChatGPT and Gemini for language polishing and for assistance with implementing experimental code; all algorithmic designs, theorems, proofs, and analyses are the authors' original work, and AI tools were not used to generate scientific ideas, hypotheses, or conclusions.

\bibliographystyle{IEEEtran}
\bibliography{sample-base}

\end{document}